\documentclass[journal,twoside,web]{ieeecolor}
\usepackage{etoolbox}
\makeatletter
\@ifundefined{color@begingroup}%
{\let\color@begingroup\relax
\let\color@endgroup\relax}{}%
\def\fix@ieeecolor@hbox#1{%
\hbox{\color@begingroup#1\color@endgroup}}
\patchcmd\@makecaption{\hbox}{\fix@ieeecolor@hbox}{}{\FAILED}
\patchcmd\@makecaption{\hbox}{\fix@ieeecolor@hbox}{}{\FAILED}

\usepackage{generic}
\usepackage{cite}
\usepackage{threeparttable}
\usepackage{amsmath,amssymb,amsfonts}
\usepackage{hyperref}
\usepackage{graphicx}
\usepackage{textcomp}
\usepackage{algorithm}
\usepackage{algpseudocode}
\usepackage{caption}

\def\BibTeX{{\rm B\kern-.05em{\sc i\kern-.025em b}\kern-.08em
    T\kern-.1667em\lower.7ex\hbox{E}\kern-.125emX}}
\usepackage{graphicx}
\usepackage{multirow}
\usepackage{array}
\begin{document}

\title{Glioma Multimodal MRI Analysis System for Tumor Layered Diagnosis via Multi-task Semi-supervised Learning}

\author{ 
Yihao Liu\thanks{Y. Liu and M. Wu are with the Cancer Research Institute, Central South University, Changsha 410083, China.},
Zhihao Cui\thanks{Z. Cui is with School of Physics \& Electronic Science, Changsha University of Science \& Technology, Changsha 410114, Hunan, China.},
Liming Li\thanks{L. Li is with the IFLYTEK Research, Anhui 230088, Hefei, China},
Junjie You\thanks{J. You is with the School of Life Sciences, Central South University, Changsha 410083, China.},
Xinle Feng\thanks{X. Feng is with the Department of Radiology, Xiangya Hospital, Central South University, Changsha 410008, Hunan, China.},\\
Jianxin Wang,~\IEEEmembership{Senior Member,~IEEE}\thanks{J. Wang is with School of Computer Science and Engineering, Central South University, Changsha 410083, China.},
Xiangyu Wang$^*$\thanks{X. Wang and Q. Liu are with the Department of Neurosurgery, Xiangya Hospital, Central South University, Changsha 410008, Hunan, China.},
Qing Liu$^*$,
Minghua Wu$^*$
\thanks{$^*$ X. Wang, Q. Liu, and M. Wu are the corresponding authors (wxymgh1989@csu.edu.cn, liuqingdr@csu.edu.cn, wuminghua554@aliyun.com).}}
\maketitle

\begin{abstract}
Gliomas are the most common primary tumors of the central nervous system. Multimodal MRI is widely used for the preliminary screening of gliomas and plays a crucial role in auxiliary diagnosis, therapeutic efficacy, and prognostic evaluation. Currently, the computer-aided diagnostic studies of gliomas using MRI have focused on independent analysis events such as tumor segmentation, grading, and radiogenomic classification, without studying inter-dependencies among these events. In this study, we propose a Glioma Multimodal MRI Analysis System (GMMAS) that utilizes a deep learning network for processing multiple events simultaneously, leveraging their inter-dependencies through an uncertainty-based multi-task learning architecture and synchronously outputting tumor region segmentation, glioma histological subtype, IDH mutation genotype, and 1p/19q chromosome disorder status. Compared with the reported single-task analysis models, GMMAS improves the precision across tumor layered diagnostic tasks. Additionally, we have employed a two-stage semi-supervised learning method, enhancing model performance by fully exploiting both labeled and unlabeled MRI samples. Further, by utilizing an adaptation module based on knowledge self-distillation and contrastive learning for cross-modal feature extraction, GMMAS exhibited robustness in situations of modality absence and revealed the differing significance of each MRI modal. Finally, based on the analysis outputs of the GMMAS, we created a visual and user-friendly platform for doctors and patients, introducing GMMAS-GPT to generate personalized prognosis evaluations and suggestions.
\end{abstract}

\vspace{0.5em}

\section{Introduction}
\label{sec:introduction}
\IEEEPARstart{G}{liomas} are the most common primary tumors of the central nervous system (CNS), accounting for ~75\% of all primary malignant brain tumors in adults. The World Health Organization (WHO) classifies gliomas as low-grade (LGGs; including WHO grades I and II, which are slow growing) and high-grade gliomas (HGGs; including WHO grades III and IV, which are fast-growing and aggressive)~\cite{louis20162016}. As the most common primary brain tumor, gliomas of varying grades have significantly different prognoses and correspond to diverse clinical treatment strategies. In 2021, WHO released its latest classification of gliomas, which now incorporates histological and molecular changes to further divide adult gliomas into distinct subtypes and introduces a layered approach to glioma diagnosis based on these characteristics~\cite{louis20212021}.

Specific genetic and molecular markers, which include mutations in the isocitrate dehydrogenase (IDH) enzyme, have become pivotal~\cite{yang2022glioma}. IDH plays a critical role in cellular metabolism, and its mutations are commonly observed in various glioma subtypes. The detection of IDH mutations has significant clinical value because it marks a subset of gliomas with distinctive prognostic features and therapeutic responses. Patients with gliomas that exhibit IDH mutations experience more favorable outcomes, including better responses to certain therapies and longer survival~\cite{lv2024insight}. In addition to IDH, this research analyzed two other significant biomarkers: 1p/19q chromosome codeletion and O6-methylguanine-DNA methyltransferase (MGMT). These genetic alterations are clinically significant as prognostic markers, which indicate various treatment responses~\cite{pirozzi2021implications, lv2024predictive}.

In the clinical landscape, the diagnosis of glioma relies on two methods: surgical interventions and imaging modalities. Imaging techniques are increasingly favored because of their cost-effectiveness, lower risk profile, and efficiency, proving indispensable not only for initial disease diagnosis but also for monitoring the course of treatment. Moreover, given the spatiotemporal heterogeneity that characterizes glioma development, MRI can provide a comprehensive three-dimensional view of the whole tumor architecture. This holistic view is crucial for capturing the complex biological behavior of gliomas, which can vary significantly across different regions of the tumor~\cite{liu2025survey}. 

Deep learning integration has been developed for enhancing MRI capabilities by automating the detection and quantification of tumor characteristics, which are critical to formulate a precise diagnosis. The diagnosis of CNS tumors is stratified into four levels: \textbf{1)} An integrated diagnosis that combines tissue-based histological analysis with molecular diagnostics. \textbf{2)} Histological diagnosis. \textbf{3)} Determination of the CNS WHO grade. \textbf{4)} Molecular information. This layered diagnostic approach is pivotal in guiding treatment decisions and affecting the selection of therapeutic interventions such as chemotherapy and radiation therapy. Comprehensive assessments enable healthcare professionals to tailor treatment plans based on the specific characteristics of a patient’s condition, optimize the chances of successful outcomes, and minimize unnecessary or ineffective interventions.

Deep learning-based methods that can learn the abstract and composite feature representations from input data have shown potential for performing medical image analysis tasks. Further, the molecular status can be characterized by MRI morphological features. A non-lobar location, larger proportion of enhancing tumors, multifocal distribution, and poorly defined non-enhancing margins were predictive of wild-type IDH1. In addition, the 1p/19q-co-deleted group often demonstrates mixed or restricted diffusion characteristics and more pial invasion compared to the non-codeletion group~\cite{park2018prediction}. These complex features can be extracted from a deep learning network and utilized to predict glioma molecular subtypes. In addition, different molecular events are not isolated, and they are interconnected and reflected in imaging features. Therefore, training a multitasking network that encourages feature sharing can help the network learn more comprehensive features and make a more accurate diagnosis~\cite{cheng2022fully}. Our multi-task model was designed for leveraging this trait and further improving the performance of various classification and segmentation tasks.

Past research on glioma diagnostic assistance has often faced issues such as large disparities in the number of samples across different categories, and incomplete annotations of genetic and histological information. These factors directly lead to inaccurate model results, significant prediction biases, and model overconfidence problems. These challenges underscore the need for more advanced deep-learning modeling techniques. Semi-supervised~\cite{hao2023predicting} and self-supervised learning-based methods~\cite{parmar2023generalizable} have emerged as an efficient way to alleviate these issues. The SimSiam contrastive learning framework~\cite{chen2021exploring} represents a pivotal advancement in self-supervised learning, enabling flexible and robust feature extraction without the need for labeled data and negative sample pairs. An uncertainty-based mechanism~\cite{kendall2018multi} has demonstrated the potential to address these data-related challenges by integrating uncertainty into the learning process.

In this study, we proposed a Glioma Multimodal MRI Analysis System (GMMAS) that applies multi-task semi-supervised learning techniques for analyzing glioma using various MRI modalities. GMMAS uses multi-task learning to simultaneously segment glioma regions and predict histological and molecular characteristics, significantly enriching diagnostic insights. The use of multi-task learning is pivotal because it allows the model to share features across tasks, enhancing the performance of each task because of the complementary nature of the information derived from each diagnostic level. Further, GMMAS utilizes a two-stage semi-supervised learning approach consisting of pseudo-labeling and consistency regularization to fully exploit both the labeled and unlabeled training data.

We introduce a novel architecture that leverages convolutional neural networks (CNNs) and transformer models to enhance the feature extraction and refinement of multimodal MRI data. In addition, our GMMAS, which is trained on multi-center datasets, effectively utilizes two types of uncertainty to conduct multi-task semi-supervised learning. Specifically, we optimize the learning process across multiple tasks via an aleatoric uncertainty-based task weighting method and select reliable pseudo labels by assessing both prediction confidence and epistemic uncertainty value. Moreover, to mitigate model overconfidence that undermines pseudo label selection, we calibrate GMMAS through label smoothing introduced by the Tumor-CutMix approach. 

Furthermore, we introduced a novel training pattern in the GMMAS architecture to improve the adaptation capability to MRI modality absences. The adaptation module consists of three stages, including a knowledge distillation process followed by contrastive learning and supervised fine-tuning. Ablation experiments demonstrated its effectiveness in enhancing the model robustness in diverse clinical settings. In addition, we employed various techniques in model training and post-processing for improving the segmentation and classification accuracies. We also used an additional U-net module to extract global MRI features, which were further fused with local representations based on the channel-attention mechanism. This approach ensures training of both input diversity and data integrity, alleviating the limitations of traditional convolutional operations with small receptive fields.
Overall, the GMMAS employs an automated analysis workflow from the input of MRI images to multiple outputs of tumor layered diagnosis. As an auxiliary diagnostic technology, this approach can reduce the workload of neurosurgeons and radiologists, enhancing their operational efficiency and effectiveness.

The key contributions of our work are detailed as follows:
\indent\textbf{1)} GMMAS leverages a novel multi-task learning architecture to generate tumor layered diagnosis, providing more comprehensive assistance for doctors.

\indent\textbf{2)} By combining a dual-threshold pseudo-labeling with consistency regularization, GMMAS effectively leverages unlabeled MRI samples to enhance model performance.

\indent\textbf{3)} The system employs our proposed adaptation module to learn cross-modal features, maintaining model performance despite the absence of various MRI modality combinations.

\indent\textbf{4)} A user-friendly platform is developed to visualize the analysis outputs and to integrate with GMMAS-GPT for personalized prognostic evaluations.

\begin{table*}
\centering
\caption{Label information of the multi-center datasets}
\label{table1}
\renewcommand{\arraystretch}{1.25}
\resizebox{0.75\textwidth}{!}{
\begin{tabular}{cccccc}
\hline
 & & Training set & Internal validation & External validation \\
\hline
\multicolumn{2}{c}{Total number} & 1184 & 296 & 309 \\
\hline
\multirow{3}{*}{Histological subtype} & GBM & 240 & 61 & 228 \\
 & LGG & 218 & 57 & 81 \\
 & N/A & 726 & 178 & 0 \\
\hline
\multirow{3}{*}{Biomarker-IDH} & Wild type & 129 & 30 & 249 \\
 & Mutant type & 71 & 21 & 60 \\
 & N/A & 984 & 245 & 0 \\
\hline
\multirow{3}{*}{Biomarker-1p/19q} & Co-deletion & 100 & 23 & 120 \\
 & Intact & 86 & 26 & 159 \\
 & N/A & 998 & 247 & 30 \\
\hline
\multirow{3}{*}{Biomarker-MGMT} & Unmethylated & 205 & 69 & 135 \\
 & Methylated & 240 & 60 & 105 \\
 & N/A & 739 & 167 & 69 \\
\hline
\end{tabular}
}
\vspace{-1em}
\end{table*}

\section{Related Literature}
In the field of brain tumors, deep-learning methodologies have been instrumental in facilitating both fundamental and intricate diagnostic tasks~\cite{liu2023deep}. Medical image segmentation and subtyping, which are often time-consuming and rely on multi-step pipelines, are crucial for delineating clear pathological structures and evaluating disease prognosis. Deep learning methods such as convolutional neural networks (CNNs)~\cite{yamashita2018convolutional} and end-to-end learning~\cite{mazurowski2019deep} have significantly boosted diagnostic efficiency. Unlike traditional machine learning methods such as random forest~\cite{wu2019radiomics}, support vector machine~\cite{ren2019noninvasive}, Bayesian model~\cite{corso2008efficient}, and clustering~\cite{parmar2015radiomic} that are limited by prior knowledge, deep learning models such as a dense convolutional auto-encoder used to classify glioma, pituitary tumor, and meningioma~\cite{waghere2024robust} and 3D U-Nets used for glioma segmentation~\cite{li2024multi} have provided an automatic analysis pipeline that augments the capabilities of medical professionals by streamlining these processes, minimizing subjective bias and enhancing diagnostic accuracy. 

For challenges, such as predicting gene mutations and risk stratification, deep learning offers insights beyond the scope of manual image analysis. Matsui et al.~\cite{matsui2020prediction} leveraged a ResNet architecture and utilized tumor center axial slices as inputs for determining the molecular classifications of low-grade glioma. However, the analysis based on slices overlooks the three-dimensional integrity of tumor images. To complement this, Tang et al.~\cite{tang2020deep} employed a 3D deep feature learning method to obtain deep features from voxels along with traditional clinical indices for predicting overall survival in patients with HGG. However, in medical image analysis, CNNs encounter limitations because of their local receptive fields, which hinder global context understanding, generalization across diverse datasets, and parameter efficiency. The Transformer~\cite{vaswani2017attention} is a deep learning structure that was first introduced for natural language processing (NLP) tasks and soon showed powerful abilities in diverse fields, including computer vision tasks~\cite{liu2023deep}. Transformers leverage self-attention mechanisms to capture long-range dependencies and global information, demonstrating enhanced generalization capabilities and scalable efficiency. Transformers are promising alternatives, offering significant potential for advancements in medical imaging because of their superior handling of global contextual insights and adaptable computational efficiency~\cite{gillot2022automatic}.

The convergence of AI technology and medical imaging has greatly influenced medical diagnostic systems. Further, existing studies have researched proposed AI platforms and applied them to clinical fields. Esteva et al.~\cite{esteva2017dermatologist} employed 129,450 clinical skin images to develop a CNN model that can identify skin cancer. Their research demonstrated that the diagnostic capability of well-trained AI models was comparable to that of dermatologists in classifying skin cancer. The development of the LYNA AI system~\cite{liu2019artificial} for breast cancer detection exemplifies the potential of deep learning to surpass traditional pathology in specific diagnostic tasks. In another study, Keenan et al.~\cite{keenan2022deeplensnet} developed DeepLensNet to assess cataract types and severity, which exhibited a significantly higher precision compared to the diagnostic accuracy of 14 ophthalmologists and 24 medical students. 

These studies indicate that deep-learning technology can improve the precision and accessibility of disease diagnosis. However, there is still no comprehensive AI diagnostic system or accessible platform for glioma layered diagnosis. Therefore, we introduced a fully automatic analysis system called GMMAS.

\begin{figure*}[!ht]
    \centering
    \includegraphics[width=\linewidth]{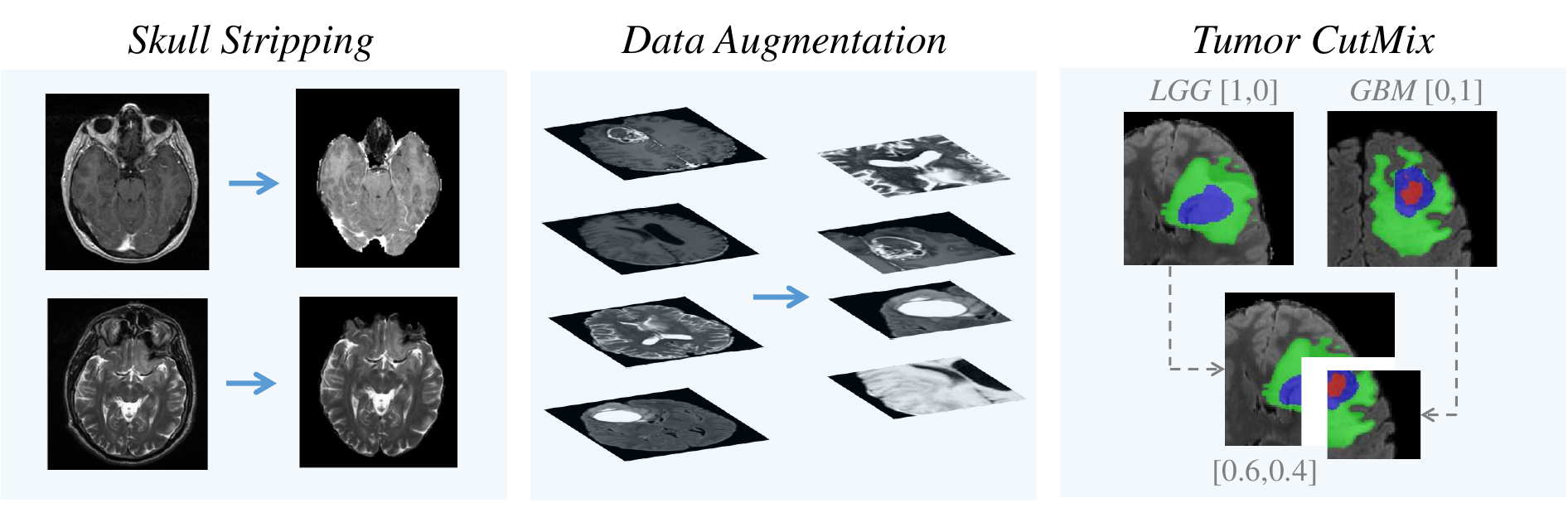}
    \caption{\textbf{Data preprocessing diagram.} Skull stripping of MRI brain images removes the skull portion, leaving only the brain tissue for analysis. Data augmentation techniques are applied to the stripped images, which involve various transformations such as cropping and flipping to increase the diversity of the dataset. ``Tumor-CutMix," a technique for creating composite images. Segments of the tumor region from one class (e.g., LGG [1,0]) are combined with tumor regions of another class (e.g., GBM [0,1]) to generate a blended tumor image, with the corresponding labels indicating the proportion of each class in the composite (e.g., [0.6,0.4]).}
    \label{fig3}
    \vspace{-1.2em}
\end{figure*}

\section{Methods}
\subsection{Datasets}
We collected image data, histological grading information, and molecular marker information from 1810 patients from four databases: BraTS 2020, BraTS 2021, TCIA, and Xiangya Hospital datasets. The collected MRI images contained six modalities: T1, T2, T1CE, T2-FLAIR, T2CE, and T2-FLAIR-CE. At the histological level, the pathological grades of 789 patients were collected, including 440 and 349 cases of glioblastoma (GBM) and low-grade glioma (LGG), respectively. We analyzed three molecular markers that are useful for gliomas to obtain molecular information. A summary of the datasets is presented in Table~\ref{table1}.

The BraTS dataset is a multicenter dataset of brain tumor segmentation challenges 2021 and 2020 (BraTS-2021 and BraTS-2020), which comprises clinically acquired multimodal MRI brain scans of patients with GBM and LGG. All ground truth labels were manually reviewed by expert board-certified neurobiologists. The data for each patient included four MRI modalities, including T1, T1CE, T2, and T2-FLAIR volumes, and the tumor ground-truth labels of segmentation data. The glioma grade in each patient was determined by a neurosurgeon and reviewed by specialized radiologists. Information on glioma-associated molecular markers for patients was obtained from The Cancer Genome Atlas (TCGA) gene database. 

In addition, we collected patient data from The Cancer Imaging Archive (TCIA) and Xiangya Hospital. The TCIA dataset is an important, publicly accessible collection of medical images for cancer research. It was established to support the cancer research community by providing a large archive of medical cancer images accessible for public download. Data in TCIA are organized into purpose-built collections typically pertaining to a specific cancer type or research focus. The collection we used from the TCIA was called LGG-1p19qDeletion. The dataset from Xiangya Hospital, which was collected by doctors, is a private dataset containing 330 patients.

In our dataset, we assumed that the label information was implied in the MRI image, and it was difficult to identify with the naked eye, including the texture, spatial shape, and boundary of the tumor. Therefore, an intelligent method is required for mining and using them.

\subsection{Data preprocessing}

As shown in Figure~\ref{fig3}, the data preprocessing workflow involves three main steps before the data can be input into a deep learning model. Each slice image in the figure represents a whole three-dimensional MRI sample. 

Step 1: ``Skull stripping" is applied to MRI images. In this process, non-brain tissues, particularly the skull, skin, and eyeballs, are removed from the image to ensure that the model focuses only on brain structures. The example shows two MRI samples from different modalities, each undergoing the skull-stripping process and resulting in clean images of the brain without the surrounding skull.

Step 2: Data augmentation~\cite{shorten2019survey} is performed. This step is critical in deep learning as it artificially expands the training dataset by creating modified versions of the input data to prevent overfitting and improve the generalization capabilities of the model. The data augmentation methods used in this study include: 1. Random Flip: Images are flipped horizontally or vertically to simulate different brain orientations, which can occur because of variations in patient positioning during the MRI scan. 2. Crop: Parts of the images are cropped randomly to create new images with different focuses, helping the model learn to identify features regardless of their position in the image. 3. Random Intensity Shift: The intensity values of the MRI images are shifted randomly to help the model become robust against variations in image brightness and contrast that may occur because of differences in MRI scanners or scanning parameters. 4. Channel Shuffle: MRI images, which are used as the original data, may have multiple channels corresponding to different MRI modalities. Shuffling these channels can teach the model not to rely on a specific order of data, making it adaptable to different multimodal MRI data configurations. Together, these data augmentation strategies significantly enrich the dataset, providing a diverse range of scenarios for the model to learn from, thereby enhancing its ability to accurately analyze new, unseen MRI images.

Step 3: Merging the MRI images through ``Tumor-CutMix." Tumor-CutMix extends the traditional CutMix approach~\cite{yun2019cutmix} to 3D glioma MRI images by implementing patch swapping and label mixing tailored for volumetric data. This method involves randomly cropping a tumor region from one MRI scan and pasting it onto another, effectively creating composite images that blend features from different glioma subtypes. Moreover, we have introduced the tumor-protect crop method, which prioritizes the preservation of tumor regions during the cropping process, ensuring that these critical areas are consistently represented in the training batches. The illustration and detailed description of this method are shown in the appendix Figure~\ref{A2}. By mixing tumor segments of MRI scans from different tumor subtypes, such as LGG and GBM, we artificially increase the diversity of training samples, enhancing the model’s ability to generalize across a wider range of tumor appearances and stages, thereby improving diagnostic accuracy. 

\begin{figure*}[t]
    \centering
    \includegraphics[width=\linewidth]{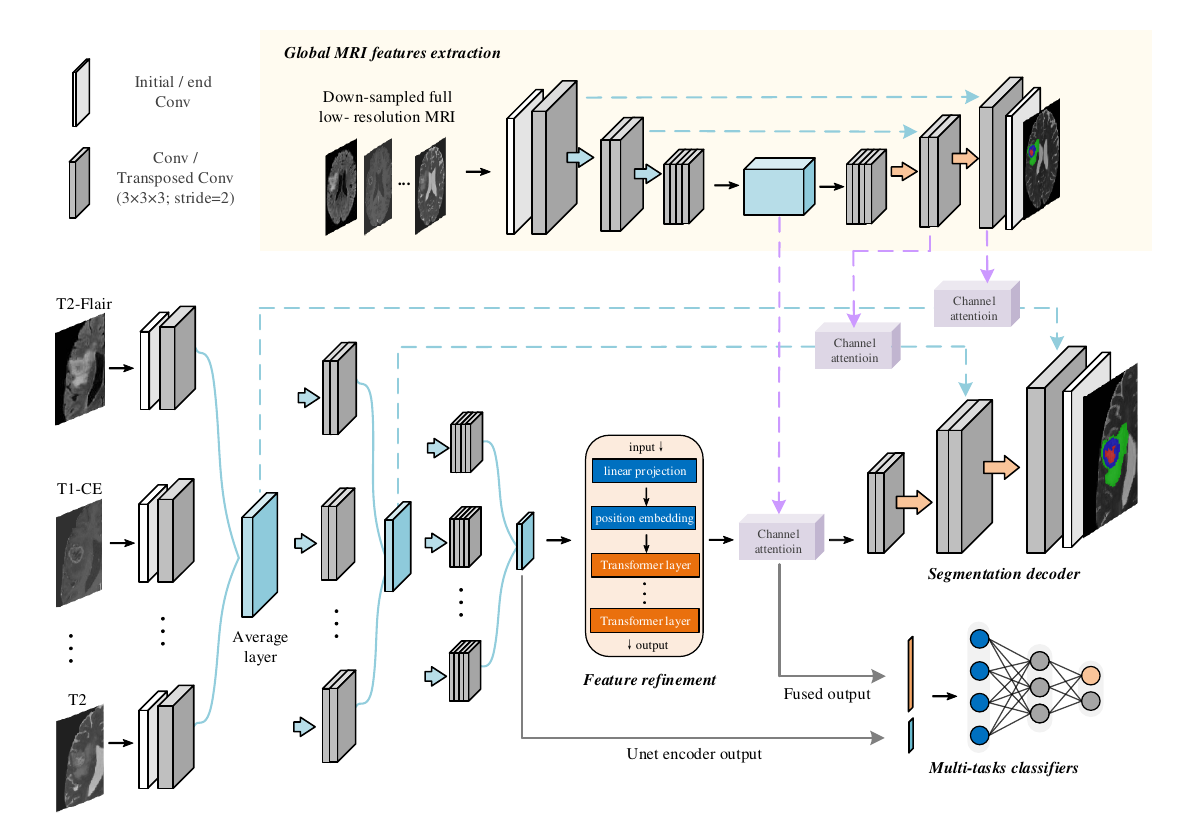}
    \caption{\textbf{Overview of GMMAS architecture.} Feature extraction begins with MRI inputs. Each modal of MRI images undergoes a transformation where features are extracted and subsequently averaged, culminating in a weighted average representation. This representation is then refined through a series of transformer layers. The refined features are passed to a segmentation decoder, which outputs the tumor segmentation on the MRI image. In parallel, the features are also fed into a multi-task classifier for glioma histological and molecular subtyping tasks. In addition, a U-net with down-sampled full MRI images as inputs was used to extract global image features to alleviate the limitations of the original patch learning. Global features of different abstract layers were merged with backbone outputs using a channel attention-based feature fusion module, as shown below. GAP: global average pooling, GMP: global max pooling.}
    \label{fig2}
    \vspace{-1.2em}
\end{figure*}

Additionally, Tumor-CutMix mitigates class imbalance by creating composite images that incorporate features from both underrepresented and overrepresented classes, ensuring a more balanced training distribution and reducing model bias toward more frequent classes. By embedding tumor regions from one subtype into the anatomical context of another, the model learns to recognize tumors in varying anatomical settings and with mixed visual features, which is crucial for accurate glioma diagnosis where tumor appearance can significantly vary based on location and surrounding brain structures. 

Furthermore, Tumor-CutMix introduces label smoothing by proportionally mixing the labels of the composite images according to the area of each tumor type present, reducing the model’s tendency toward overconfidence in its predictions and promoting a more generalized decision-making process. This augmentation strategy not only enhances model robustness against overfitting but also improves sensitivity and specificity by exposing the model to complex, mixed patterns not typically present in standard training datasets.

\subsection{GMMAS model development}
In accordance with our clinical requirements, a deep learning network was designed comprising a multi-input CNN encoder for feature extraction, a transformer encoder for feature refinement, and two parallel architectures, including a decoder for glioma segmentation and a classifier for glioma grading, IDH genotyping, 1p/19q chromosome disorder, and MGMT status prediction. This architecture (GMMAS) is shown in Figure~\ref{fig2} and consists of four main components: 

\textbf{1) CNN-Transformer Encoder Architecture:} This architecture leverages the synergy between CNNs and the multihead self-attention mechanism of the transformer to address the limitations of CNNs in capturing long-range dependencies and streamlining computational efficiency. The feature-acquisition process can be divided into two stages.

The first stage is feature extraction, wherein convolutional layers are used to process the input data from different modalities and extract the corresponding features. The input tensor was $x\in\mathbb{R}^{C\times H\times W\times D}$, where $H\times W$, $D$, and $C$ represent the spatial resolution, depth dimension (number of slices) of the MRI samples, and number of channels (referred to as the modality number), respectively. Initially, the input was convoluted with a   kernel, producing a 16-channel feature map for each MRI modal. This structure decouples the MRI inputs of different modalities and fuses the information using the weighted average of the feature maps. Considering that modalities may have different significance for layered diagnosis tasks, we set the weights of each feature map as learnable parameters during model training. This makes it possible to train a robust model for modal absences. The feature map $F_{merged}$ can be expressed as:
\begin{equation}
    \begin{aligned}
     & F_{merged}=(W_{flair}\times F_{flair}+W_{Tlce}\times F_{Tlce}+ \\
     & W_{Tl}\times F_{Tl} + W_{T2}\times F_{T2})/W_{sum}
    \end{aligned}
\end{equation}
where $W$ and $F$ represent the weight parameters of each modal and output feature maps of the first convolution layer, respectively. Subsequently, the merged features are added to the up-sampling layers of the corresponding segmentation decoder to share valuable information and construct a U-net architecture, which is a classic segmentation network~\cite{falk2019u}. The merged feature map was subjected to three down-sampling steps and layers of residual convolution blocks to further extract the spatial and depth features. Each down-sampling stage employs a   convolution with a stride of 2, doubling the channel count of the output relative to the input. The Resnet block, which contains two convolution layers with a residual connection, utilizes batch normalization, Leaky ReLU activation, and a $3\times 3\times 3$ convolution, culminating in a feature map $F_\lambda\in\mathbb{R}^{2^{\lambda+3}\times(H/2^{\lambda-1})\times(W/2^{\lambda-1})\times(D/2^{\lambda-1})}$ at the $\lambda$ layer of the feature extraction stage.

The encoder employs Transformers to refine the feature representation and capture the global semantic features of MRI inputs, extending the vision transformer’s (ViT) approach~\cite{han2022survey} of splitting images into patches. The volumetric feature map ($\lambda$ = 4) of the merged MRI modals is passed to feature refinement stage as 3D patches ($16\times 16\times 16$). During linear projection, to transform the feature space and enhance the representation of every patch, all 3D patches undergo a $3\times 3\times 3$ convolution with a stride of 1 increasing channel from 128 ($2^\lambda+3$) to 512, and they are then flattened into $N_{embed}\times d$ tokens and combined with a positional embedding to maintain spatial information, where $N_embed$ represents the number of embedding channels equal to 512 and d represents the number of flattened patches calculated by $16\times 16\times 16$. The Transformer processes these tokens through four layers built on a multihead self-attention (MHSA) module, followed by a multilayer perceptron (MLP). In the computation of self-attention, the matrices Q (query), K (key), and V (value) are utilized with their own linear projection parameters, $W_{Q}, W_{K}, W_{V}$, which result in $\mathrm{SA~Input}=\mathrm{F}(W_{_O},W_{_K},W_{_V},\text{feature embedding})$. The presence of the MHSA enables the network to focus on different spatial positions and diverse feature subspaces, which can be formulated as:
\begin{equation}\mathrm{MHSA}=\text{Linear}(\mathrm{Concat}(SA_1,SA_2,SA_3,SA_4))\end{equation}

\textbf{2) Decoder for Glioma Segmentation:} This decoder leverages cascaded up-sampling layers to transform high-level feature maps back to their original resolution for voxel-level glioma segmentation. After converting the feature sequence to a 3D format, successive up-sampling steps, each comprising a $1\times 1\times 1$ convolution for dimension reduction and a $3\times 3\times 3$ convolution for upscaling, were applied. Skip connections merge down-sampled and up-sampled features and are finalized using a softmax function for segmentation to enrich the segmentation with spatial details.

\textbf{3) Multi-task Classifier:} A multiscale CNN classifier employs encoder-generator feature maps for multi-task classification. In our model, the classification module receives two types of input: one from the U-Net encoder and another from the transformer combined with an additional global feature fusion module. The U-Net encoder processes each MRI modality through multiple convolutional layers, progressively downscaling and extracting features at multiple levels of abstraction. On the other hand, the transformer refines the features through self-attention mechanisms, exploiting global contextual information. This is further enhanced by fusion with features learned by the additional global feature extraction module, generating a comprehensive feature map that supports deeper contextual understanding. The integration of these two types of feature inputs enables the model to handle complex classifications by leveraging both detailed anatomical information and global contextual relationships. The concatenated features were then processed through ``Summarize and Separate" blocks to consolidate information before branching into separate interconnected prediction tasks, culminating in a softmax output for multiple tasks including glioma grading, IDH mutation status prediction, 1p/19q chromosome disorder prediction, and MGMT status prediction. The detailed architecture design of the multi-task Classifier is shown in the appendix Figure~\ref{A1}.

\textbf{4) Global feature fusion:} We utilized a U-net architecture with encoders and decoders adapted from the backbone network to harness global image features from full but down-sampled MRI images. The architecture was fine-tuned using down-sampled segmentation labels under supervised learning, enabling the extraction of comprehensive global features crucial for detailed image analysis. These global features were integrated with the backbone outputs using a channel attention-based feature fusion module. This module begins with the concatenation of feature maps followed by both global average pooling and global max pooling to capture different statistical perspectives of the features. The attention-enhanced output is recalibrated using a convolutional layer to restore it to the original dimensions of the feature space.

\subsection{Multi-task learning loss functions}
We implemented a multi-task learning method based on uncertainty to manage samples with various labels. For instances lacking certain labels, the model makes dynamic adjustments by excluding these instances from the multi-task loss computation, directing the optimization process in a more focused manner. Hence, the design of an appropriate multi-task loss function is of great significance for this model.

For glioma segmentation, a crucial metric to evaluate the performance of a segmentation model is the Dice coefficient, which measures the similarity between the predicted segmentation and ground truth. The Dice loss, which is utilized as a loss function during the training of neural networks for segmentation tasks, is designed to optimize the Dice coefficient. The Dice coefficient is defined as twice the area of overlap between the predicted segmentation and ground truth, divided by the total number of pixels in both the predicted and ground truth segmentation. The Dice coefficient is formulated as:
\begin{equation}DiceLoss=1-\frac{2\times|Y\cap\hat{Y}|}{|Y|+|\hat{Y}|}\end{equation}
where $Y$ and $\hat{Y}$ represent the ground truth binary segmentation map and predicted binary segmentation map, respectively. Based on the Dice coefficient, the loss function was designed for ensuring that the loss was minimized when the overlap between the predicted segmentation and ground truth is maximized, encouraging the model to produce segmentation that closely match the ground truth. The Dice-based loss function can be formulated as:
\begin{equation}L_{seg}=1-\frac{2}{N_{SegClass}}\sum_{j\in N}\frac{\sum\hat{y}_j^{(i)}y_j^{(i)}}{\sum\hat{y}_j^{(i)}+\sum y_j^{(i)}+\epsilon}\end{equation}
where $y_j^{(i)}$ and $\hat{y}_j^{(i)}$ represent the ground truth segmentation labels and model softmax outputs of tumor subregion class $j$, respectively, and $N_{SegClass}$ represents the total segmentation class number, which refers to four in this research (tumor necrosis region, enhancing tumor region, edema region, and non-tumor region). Further, $\epsilon$ represents a tiny constant used in practice to prevent a zero denominator.

\begin{figure*}[t]
    \centering
    \includegraphics[width=\linewidth]{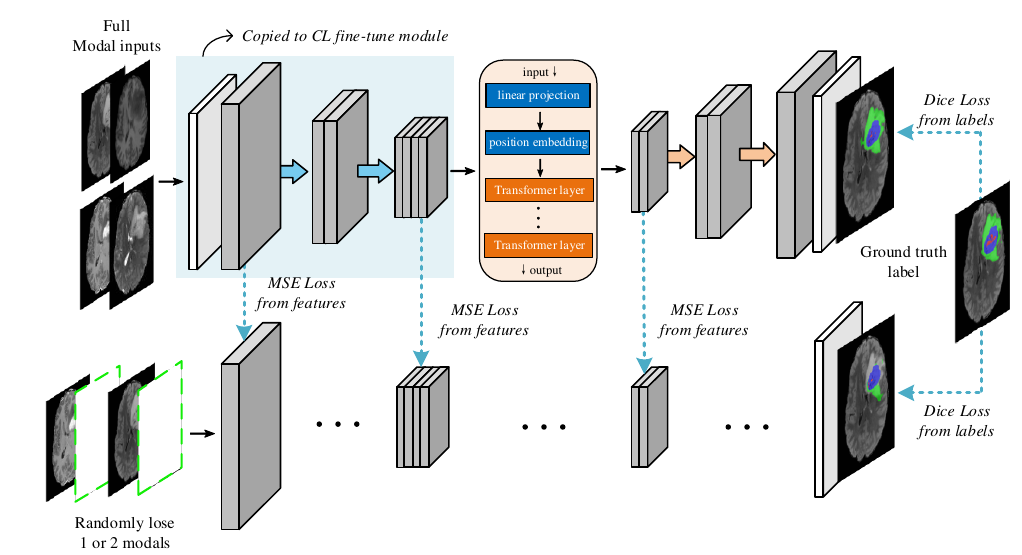}
    \caption{\textbf{Illustration of the first stage of adaptation module.} A dual-pathway knowledge self-distillation structure. The top pathway processes full-modal inputs, while the bottom pathway handles cases where one or two modalities are randomly omitted, simulating scenarios with incomplete data. Mean squared error (MSE) loss from features between pathways are used to help the model learn cross-modal features and maintain performance. Dice losses from the ground truth labels are used to refine the segmentation accuracy of every pathway. (b) A contrastive learning Siamese network for cross-modal feature extraction in MRI Images.}
    \label{fig4_1}
    \vspace{-1.2em}
\end{figure*}

The imbalanced distribution of data across different subtypes significantly affects the glioma histological and molecular subtyping performance of the model. To alleviate this issue, weighted cross-entropy loss considers the class imbalance in the dataset by assigning different weights to different classes, helping the model assign more importance to the minority class during training. For a classification problem with $C$ classes ($C$ = 2 in this study), each class has its own weight factor. The predicted probabilities for each class are denoted by $P_i$, and the ground-truth one-hot encoded labels are denoted by $y_i$ for class $i$. The weight of class $i$ is denoted as $w_i$. The weighted cross-entropy loss for a single sample is given by:
\begin{equation}L_{CrossEntropy}=-\sum_{i=1}^cw_i\cdot y_i\cdot\log(P_i)\end{equation}
$w_i$ can be calculated as $n_i/\sum_{i=1}^Cn_i$, where $n_i$ represents the number of samples of class $i$. In this way, different types of samples are assigned different weights, and the model focuses on fewer categories and helps balance the training process.

Furthermore, we adopted an uncertainty-based strategy~\cite{kendall2018multi} to train multiple tasks simultaneously for learning different information from MRI images and mitigating the potential negative consequences of multi-task learning, which introduces the risk of task bias when the weighting assigned to different tasks is not calibrated correctly. Further, they divided the uncertainty in data modeling into two types to achieve a balance between every task while learning information from the inputs: epistemic and aleatoric uncertainty.

Epistemic uncertainty refers to the uncertainty inherent in a model because of limited data. This type of uncertainty can be reduced by collecting more data or improving the architecture of the model and the training methods. The epistemic uncertainty is assessed using the MC-Dropout method~\cite{kendall2017uncertainties}, which can be mathematically described by the standard deviation of the predictions from multiple stochastic forward passes. For a given input $x$, the uncertainty $U_\mathrm{epistemic}(x)$ can be calculated as:
\begin{equation}
\label{equ6}
U_\mathrm{epistemic}(x)=\sqrt{\frac{1}{M}\sum_{m=1}^M\left(f^m(x)-\overline{f}(x)\right)^2}
\end{equation}
where $f^m(x)$ represents the $m^th$ stochastic forward pass out of $M$ total passes and $\overline{f}(x)$ represents the mean prediction across all passes. The epistemic uncertainty values computed by MC-Dropout are applied during the inference phase after training. The experimental results in Section 4.4 show that there is a strong correlation between the epistemic uncertainty values and the model calibration, which refers to the degree to which the probabilities predicted by the model reflect the true likelihood of outcomes. Therefore, based on the epistemic uncertainty threshold, we can filter and select the more reliable predictions to generate high-quality pseudo-labels to perform semi-supervised learning in this study.

Aleatoric uncertainty represents the inherent noise and variability in the observational data that cannot be reduced by obtaining more data. Task-specific aleatoric uncertainty captures the inherent noise and unpredictability in the data for each task. The model can adjust the importance of the loss of each task based on the variability inherent in that task by quantifying this uncertainty. This enables us to leverage the uncertainty values to balance weights between different tasks. The key idea is to introduce a measure of uncertainty for each task, which can be used to weight the loss of each task in the overall optimization process, ensuring that each task contributes appropriately to the model based on its specific characteristics and challenges. For the $i^th$ task, $\sigma_i$ is defined as its aleatoric uncertainty value, which is a trainable parameter. The general formula for calculating the task-weighted loss $L_{total}$ in a multi-task learning framework can be expressed as:
\begin{equation}L_{total}=\sum_{i=1}^{N_{task}}\frac{1}{2\sigma_i^2}L_i+\log\sigma_i\end{equation}
where $L_i$ represents the loss value of each task to be minimized during training and $N_{task}$ represents the number of tasks. Further, $\sigma_i$ represents the estimated uncertainty of the $i^th$ task, which acts as a weighting factor. The term $(1/2\sigma_i^2)L_i$ adjusts the contribution of the loss of each task based on its uncertainty, and $\log\sigma_i$ represents a regularization term that prevents the uncertainty from becoming too large, which would trivialize the loss. This approach avoids overfitting to tasks with more noise and instead focuses on tasks with higher data quality and lower uncertainty. Tasks with higher uncertainty have their losses down-weighted, preventing them from dominating the learning process when their prediction confidence is low. In practice, aleatoric uncertainty values are set as trainable parameters. In this study, the multi-task loss functions include $\sigma_{seg}$, $\sigma_{idh}$, $\sigma_{mgmt}$, $\sigma_{lpl9q}$, and $\sigma_{grade}$, which are the uncertainty weights and learnable parameters for network learning.
\begin{equation}L_{joint}=\frac{1}{2\sigma_{seg}^2}L_{seg}+\frac{1}{2\sigma_{cls}^2}L_{cls}+\log(\sigma_{seg}\sigma_{cls})\end{equation}

\begin{equation}
\begin{aligned}
& L_{cls}=\frac{1}{2\sigma_{idh}^2}L_{idh}+\frac{1}{2\sigma_{mgmt}^2}L_{mgmt}+\frac{1}{2\sigma_{grade}^2}L_{grade}\\
& +\frac{1}{2\sigma_{lpl9q}^2}L_{lpl9q}+\log(\sigma_{idh}\sigma_{mgmt}\sigma_{grade}\sigma_{lpl9q})
\end{aligned}
\end{equation}

\begin{figure*}[t]
    \centering
    \includegraphics[width=\linewidth]{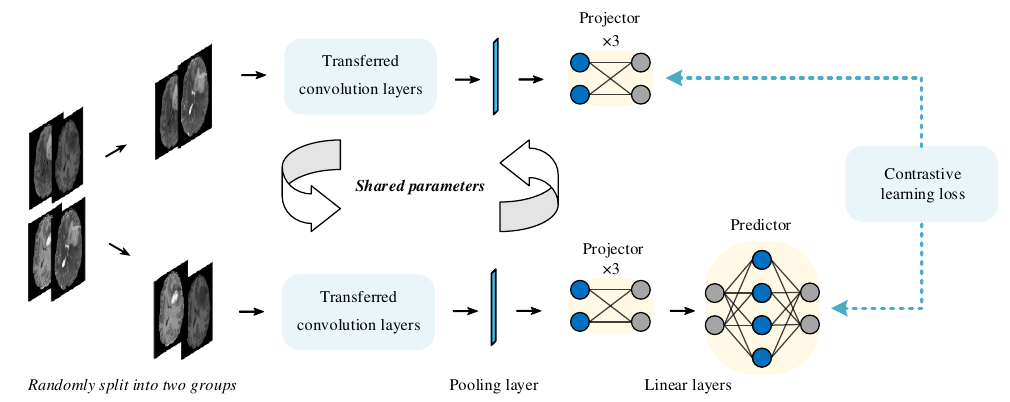}
    \caption{\textbf{Illustration of the second stage of adaptation module.} A contrastive learning Siamese network for cross-modal feature extraction in MRI Images.}
    \label{fig4_2}
    \vspace{-1.2em}
\end{figure*}

\subsection{Adaptation module for cross-modal feature learning}
Missing MRI modalities is a common occurrence in clinical practice~\cite{sharma2019missing}. The absence of one or more modalities in a patient dataset can significantly hamper the utility of deep learning models because of the resultant loss of critical information, which can affect the performance and robustness of the models. Consequently, the inability of a model to adapt to missing modalities while maintaining high performance limits the effective use of diverse clinical samples and poses a substantial barrier to integrating such models into clinical workflows. Therefore, devising a method that enables the model to adapt to any configuration of missing modalities is imperative.

Cross-modal features include attributes and patterns that are identifiable across multiple data modalities, transcending the limitations of individual modal types. Such features offer significant generalizability and robustness, which are essential for building models that can effectively adapt to missing or incomplete data sets. In this study, we designed an adaptation module which learns cross-modal feature patterns to help GMMAS extract more universal MRI features and thus improve its robustness under different modality absences.

The entire adaptation module consists of two stages, including Stage 1: Knowledge Self-Distillation and Stage 2: Contrastive Learning Fine Tuning. Knowledge self-distillation~\cite{zhang2021self} is a learning strategy where a simpler ``student" model improves its performance by mimicking the output of a more complex ``teacher" model. In this study, we employ a knowledge self-distillation approach to train the network to cope with missing modalities, as shown in Figure~\ref{fig4_1}. We leverage a ``teacher" model with access to all modalities to guide a ``student" model that sees only a subset of the modalities. This method allows the student model to learn representations that the teacher model has learned from the full set of modalities, enabling it to make accurate predictions, even when some modalities are missing.

Specifically, our architecture employs a learning mechanism where the model parameters are optimized during training to enhance its ability to adapt when certain modality inputs are absent. We have introduced a dual-pathway training strategy to enable our model to learn hidden cross-modal features from the inputs:

\textbf{I)} Full Modality Pathway: All modality inputs are present, and the model learns to optimize its parameters in a standard multi-modal learning environment. \textbf{II)} Modality Absence Pathway: One or more modality inputs are randomly excluded during training. The model is trained to predict the missing modality features by minimizing the Mean Squared Error (MSE) between the features produced with all modalities and those produced with missing modalities. The MSE loss can be formulated as:
\begin{equation}\mathrm{MSE~Loss}=\frac{1}{N_e}\sum_{i=1}^{N_e}(F_\mathrm{full~}-F_\mathrm{partial})^2\end{equation}
where $F_{full}$ represents the feature maps from the network with full modalities and $F_{partial}$ represents the feature maps from the network with incomplete modalities. Ne represents the total number of elements in the feature map. These losses are from different levels of feature maps and form a deep supervision~\cite{luo2021learning} to stimulate the learning performance. Deep supervision is incorporated at critical points to ensure robust feature learning, including the first and last convolutional layers of the encoder and the first convolutional layer of the decoder. In the meantime, the model also gets supervision signals from segmentation labels to optimize the network. This dual-objective approach ensures that the model retains its performance and generalizes well across various modality combinations.

Figure~\ref{fig4_2} illustrates stage 2, which is a Siamese network architecture, specifically designed for contrastive learning in the context of MRI image analysis. The SimSiam-based network is divided into two branches sharing the same parameters to ensure a uniform feature extraction process. Each branch processes a subset of the data that has been randomly split into two groups containing various MRI modalities. Both branches utilize transferred convolutional layers copied from stage 1 to ensure consistent feature learning. After processing through these layers, the data passes through a pooling layer, which reduces dimensionality and prepares the features for subsequent linear parts.

\begin{figure*}[t]
    \centering
    \includegraphics[width=\linewidth]{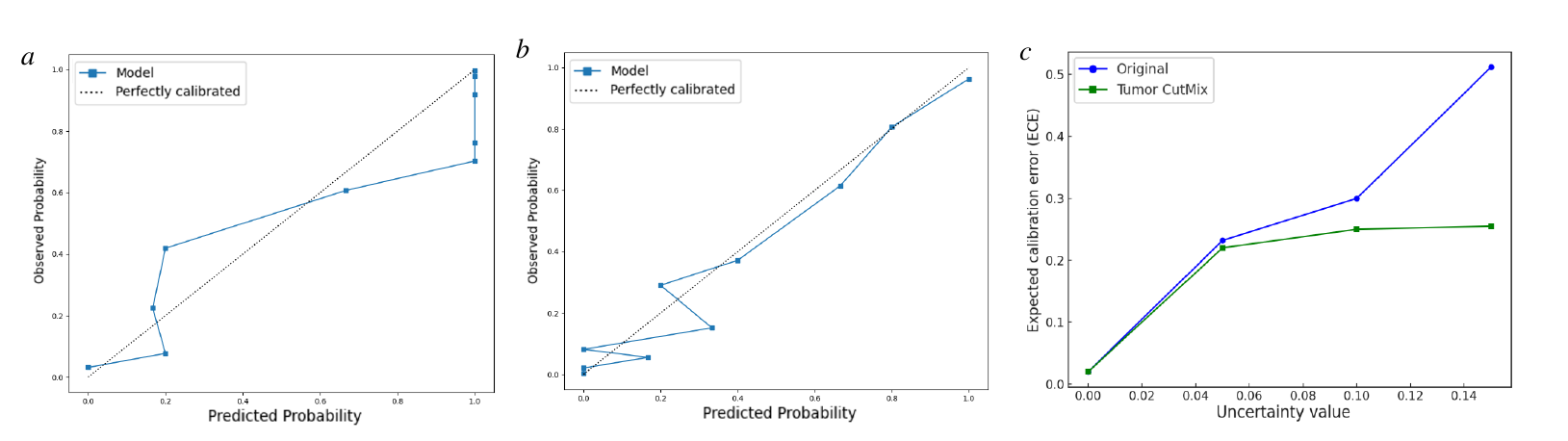}
    \caption{Calibration curves (a) before and (b) after introducing the (c) Tumor-CutMix strategy ECE values versus prediction epistemic uncertainty values of the uncalibrated and calibrated models.}
    \label{fig9}
\end{figure*}

Each branch then feeds into a projector consisting of three linear layers, which are designed to refine and project the features into a space conducive to contrastive comparison. One branch extends into a predictor, and the contrastive learning loss is computed using the output from the predictor of one branch and the output from the projector of the other branch. Specifically, the Contrastive Learning (CL) loss function can be expressed as follows:
\begin{equation}\textit{CL Loss}=-\frac{1}{2}\left(\cos\left(z_1,p_2\right)+\cos\left(z_2,p_1\right)\right)\end{equation}
where $z_1$ and $z_2$ are the outputs of the projector, $p_1$ and $p_2$ are the outputs of the predictor, and cos represents the calculation of cosine similarity. This symmetrical loss calculation ensures that the learned representations are consistent across different groups of MRI modal pairs, enhancing the model's ability to generalize from one view to another by maximizing the cosine similarity between corresponding projector and predictor outputs. In our study, by minimizing the distance between different MRI modal pairs from the same sample, it helps to refine the network's ability to discriminate and align features across modalities. This approach ensures that the network learns a feature space where the same entities across different modalities are represented similarly, thus effectively capturing cross-modal features.

It's noteworthy that MSE loss and Cosine Similarity loss emphasize different aspects of similarity: MSE loss prioritizes numerical values or magnitudes, making it sensitive to the scale of data, while CL loss focuses on the direction of the data vectors, independent of their magnitude. Combining these two types of losses in different stages allows the model to better learn cross-modal features, capturing both magnitude and directionality, which is crucial for robust representation learning. The adaptation module is followed by a supervised fine-tuning process involving both segmentation and classification branches with unchanged convolutional encoder parameters to reserve the ability to extract universal features. These three steps constitute our adaptation module in GMMAS to face different modality combination inputs.

\subsection{Model calibration and Semi-supervised learning}
Model calibration—an essential aspect of trustworthy AI systems—refers to the alignment of the confidence levels of models with their actual prediction accuracy~\cite{van2016calibration}. An ideally calibrated model produces confidence scores that accurately reflect the likelihood of making correct predictions. However, in practice, models tend to exhibit miscalibration, manifesting as either underconfidence or overconfidence in their predictions. This discrepancy between confidence and accuracy can impair the utility of the model in critical decision-making scenarios where reliability is paramount, such as in clinical diagnoses.

\begin{figure*}[t]
    \centering
    \includegraphics[width=\linewidth]{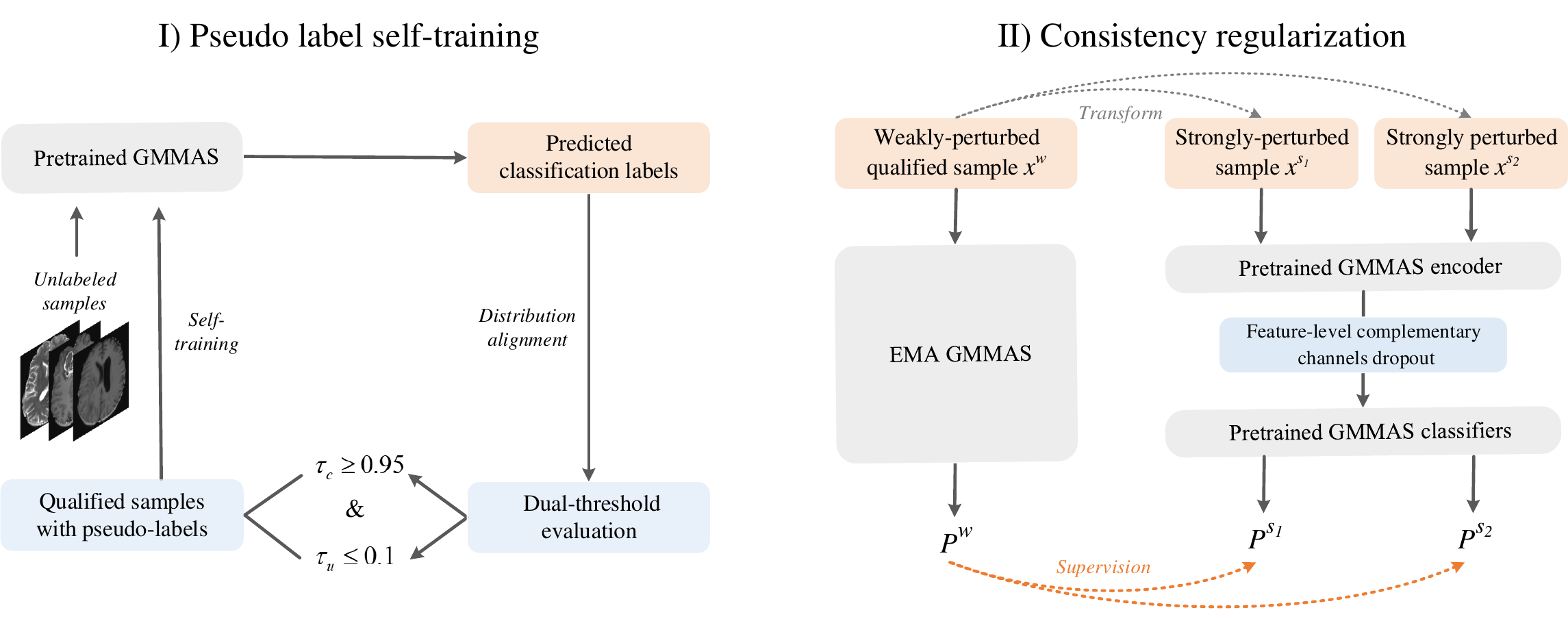}
    \caption{\textbf{Illustration of our two-stage semi-supervised learning method.} In the first stage, a dual-threshold evaluation strategy is utilized to select reliable pseudo classification labels, which are then used in the self-training process and input into the second stage. In the second stage, the weak-to-strong consistency regularization approach is utilized to fully exploit the effectiveness of the semi-supervised learning.}
    \label{fig10}
\end{figure*}

Model overconfidence is a prevalent issue in deep learning models and arises from several factors, including data imbalance, limited diversity in the training set, and intrinsic complexity of the model. Data imbalance, in which certain classes are significantly overrepresented compared with others, can lead models to exhibit a bias towards these classes and falsely inflate confidence in their predictions. Additionally, a lack of diversity in the training data can also lead the models to overfit the limited training examples and generalize poorly to unseen data. Furthermore, this overfitting can be exacerbated in models with high complexity or capacity, which, despite their ability to capture intricate patterns in the data, can become overly confident in their predictions because they memorize rather than learn generalizable features. 

In this study, GMMAS adopted the Tumor-CutMix strategy in the supervised training phase to mitigate this challenge. We aimed to achieve a more calibrated and generalizable model by leveraging the ability of CutMix to temper model confidence and enrich the learning process using a diverse set of features. By introducing label ambiguity and visual diversity through Tumor-CutMix during model training, GMMAS develops a better-calibrated prediction.

Figure~\ref{fig9}(a) shows a calibration curve comparing the predicted probabilities of the uncalibrated model with the observed frequencies. The dashed line represents a perfectly calibrated model in which the predicted probabilities match the observed frequencies. The solid line with markers indicates the performance of the model under calibration degree examination. A significant deviation from the dashed line suggests that the model was poorly calibrated, with a tendency to be overconfident in its predictions, as indicated by the overestimation at higher predicted probabilities. Figure~\ref{fig9}(b) illustrates the calibration curves after adjustment using Tumor-CutMix. The curves demonstrate the calibration effect of this technique on the model predictions. Although the calibration of the original model showed a substantial deviation from the ideal calibration, the calibration curve after the adjustment moved closer to the dashed line, indicating an improvement in the calibration degree. Figure~\ref{fig9}(c) depicts the relationship between the expected calibration error (ECE) and prediction epistemic uncertainty values, which indicate the coverage of training data and are quantified by Equation~\ref{equ6} for both the original and Tumor-CutMix-adjusted models. The ECE metric measures the average absolute difference between a model's predicted confidence and the actual correctness, assessing the model's reliability. As prediction epistemic uncertainty increased, the original model's ECE rose sharply, while the Tumor-CutMix model sustained a lower ECE, suggesting enhanced calibration across uncertainty levels.

Semi-supervised learning is a machine-learning paradigm that lies between the realms of supervised and unsupervised learning. This involves training a model on a dataset in which only a portion of the data is labeled, while the remaining data are left unlabeled. In our study, there are still a large number of samples in datasets without biomarker labels because genetic information and biomarker labels are relatively costly and difficult to collect completely, significantly limiting the performance and generalization abilities of the model. Hence, we applied a semi-supervised learning method to alleviate the limitations of data label incompleteness. Pseudo-labeling is a technique in which unlabeled data points are assigned pseudo-labels based on the predictions of a model trained on an initially labeled dataset. The idea is to use these pseudo-labeled data points actively to retrain the model, improving its performance by leveraging additional data.

As shown in Figure~\ref{fig9}(c), there is an observable trend in which increased epistemic uncertainty values in the model predictions are associated with higher expected calibration errors. Therefore, we can set a threshold for the uncertainty value to filter out more reliable predictions as one of the metrics for selecting reliable pseudo-labels. Specifically, we employed an uncertainty-based pseudo-labeling method~\cite{kendall2018multi} to conduct the first stage of our semi-supervised learning, the pattern of which is shown in Figure~\ref{fig10}.

In this study, we utilized a two-stage approach for semi-supervised learning. In the first stage, we aimed to select more reliable model predictions which have lower epistemic uncertainty, presumably lower expected calibration errors, and higher confidence. Additionally, before we evaluated the predictions, we employed distribution alignment~\cite{berthelot2019remixmatch} to adjust the model’s predicted class probabilities on unlabeled examples by scaling them according to the ratio of the true class distribution to the model’s predicted distribution. The adjusted predictions were then renormalized to form a valid probability distribution and made ready for the evaluation. Our pseudo-labeling approach rigorously quantifies uncertainty using the MC-Dropout technique, as outlined by~\cite{kendall2017uncertainties}. This method involves performing multiple stochastic forward passes through the network, using dropout at test time for providing a probabilistic interpretation of the predictions of the model. The standard deviation of these predictions reflects epistemic uncertainty, quantifying variability inherent in the model attributed to limited data or model architecture. We set a stringent uncertainty threshold $\tau_u\leq0.1$ for ensuring that only predictions with minimal uncertainty are considered, reducing the potential for erroneous label inclusion. Concurrently, we employ a high-confidence threshold $\tau_c\geq0.95$, ensuring that only the most probable predictions are used for generating pseudo-labels. This dual-threshold approach guarantees that the selected pseudo-labels are highly reliable and representative of confident model predictions, enhancing the overall integrity and effectiveness of the training process. This selective strategy is shown in the left Figure~\ref{fig10}, which illustrates the flow of unlabeled data through our semi-supervised method.

The second stage leverages a weak-to-strong consistency regularization strategy to improve semi-supervised learning, as shown in the right Figure~\ref{fig10}. In this stage, GMMAS first generates pseudo-labels by making predictions on weakly-augmented versions of the MRI samples. These weak augmentations are basic transformations including cropping, flipping, and resizing. Then, the model learns by training on strongly-augmented versions of these images, which undergo more intense transformations including color jittering, CutMix, and Gaussian blurring. For color jittering in grayscale images like MRI scans, only brightness and contrast adjustments are suitable, as saturation and hue shifts are not applicable due to the absence of color information. This method forms a teacher-student setup, where the EMA (Exponential Moving Average) teacher model provides stable pseudo-labels that guide the student model with strongly-augmented images during training, enforcing robustness and invariance under more challenging conditions. The architecture is inspired by UniMatch V2~\cite{yang2025unimatch} by unifying image and feature-level augmentations into a single stream. This method applies complementary dropout techniques to feature maps, creating dual streams of augmented features for improved learning. Specifically, feature maps are split using a binomial distribution to generate two complementary dropout masks, which allow the model to process distinct feature sets for more robust learning.

\section{Results}
\subsection{Implementation details}
The GMMAS model was developed using the PyTorch framework and executed on eight NVIDIA V100 GPUs. We adopted a cross-validation approach, in which 80\% of the data (1445 samples, 79.8\%) were allocated for model training, and the remaining 20\% (365 samples, 20.2\%) were used for model performance validation and testing. The model utilizes a stochastic gradient descent (SGD) optimizer that features a gradually reducing learning rate, starting at 0.0001.

For our experiments, we configured the batch size to two per GPU and planned the training process to run for up to 500 epochs to ensure comprehensive learning. We employed the Dice similarity coefficient (Dice value) to measure the accuracy of glioma segmentation to assess and validate the performance of GMMAS and the comparative models. In addition, the effectiveness of the model in handling multiple classification tasks has been evaluated using key metrics such as the area under the receiver operating characteristic curve (AUC), accuracy (Acc), and F1 score. These metrics provide a comprehensive quantification of the model performance across various testing scenarios.

\subsection{Model performance}
\subsubsection{Tumor segmentation task}

\begin{figure}[t]
    \centering
    \includegraphics[width=\linewidth]{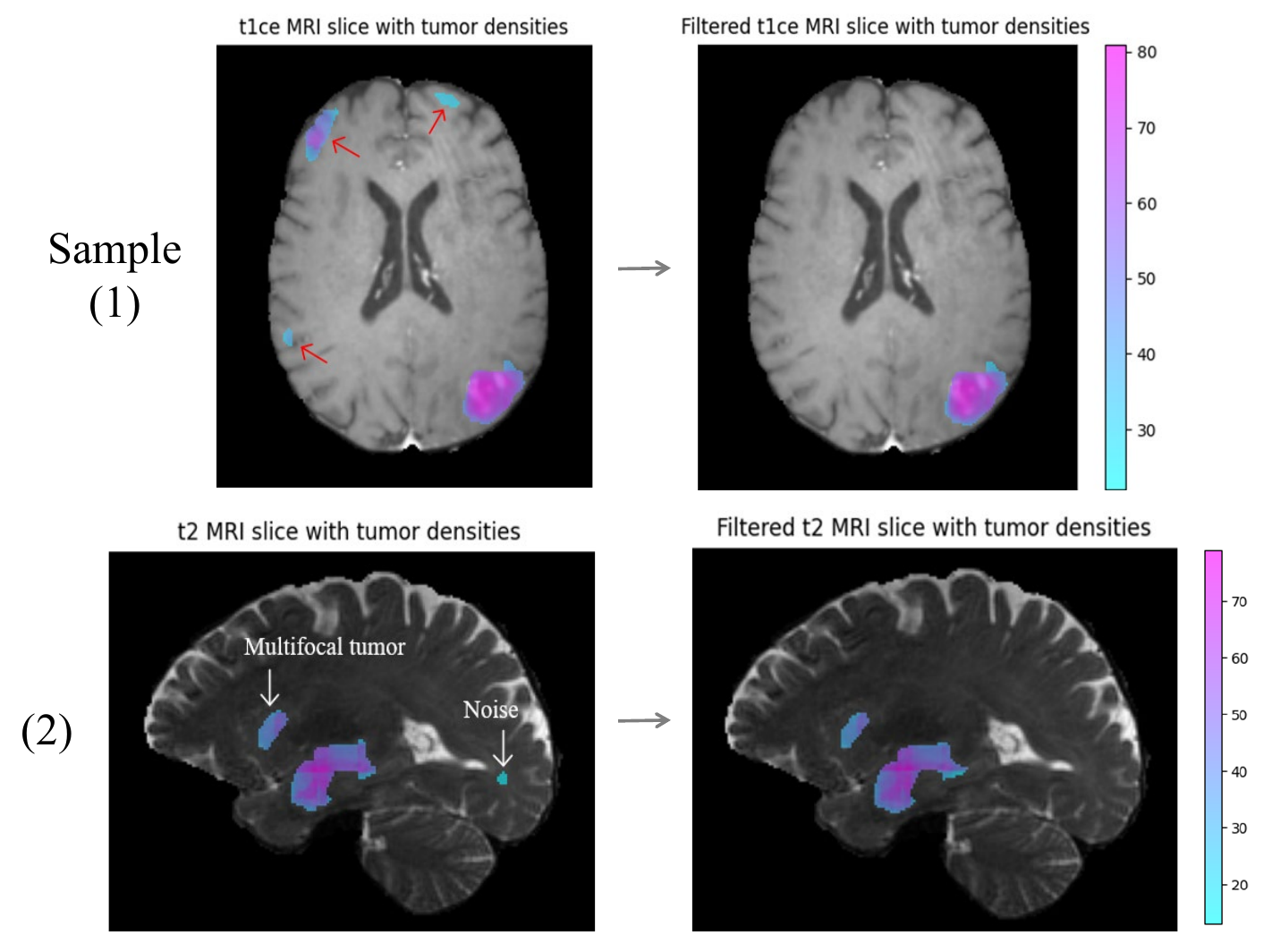}
    \caption{\textbf{Glioma MRI images before and after automatic tumor filter.} Sample (1) demonstrates a T1CE MRI slice with the highlighted predicted tumor region before and after the filtering process. Sample (2) shows a T2 MRI slice and illustrates the removal of noise and retaining a multifocal tumor following filtration.}
    \label{fig8}
    \vspace{-1em}
\end{figure}

For more precise tumor segmentation, we integrate a post-processing method to filter the falsely predicted voxels based on tumor densities, which are defined by the aggregation degree of the predicted tumor voxels. The filter process is shown in the appendix pseudocode (see Algorithm~\ref{algA1}) and Figure~\ref{fig8}.

Figure~\ref{fig8} shows the results of the automatic tumor filter applied to T1CE and T2 MRI slices for accurate delineation of tumor regions. The left images display the original MRI slices with the tumor density value highlighted, where erroneous predictions by the model are indicated by arrows in the T1CE slice of sample (1). In sample (2), the patient had a multifocal glioma. The tumor filter automatically retained the multifocal tumor region and eliminated the noise output voxels. The images on the right depict the corresponding slices after filter application, where false positives (non-tumorous voxels incorrectly classified as tumors) were reduced significantly. 

Table~\ref{table2} shows that Dice scores quantitatively measure the segmentation accuracy of our GMMAS under different enhancement strategies. These scores reflect the system's ability to delineate three key tumor regions: the whole tumor, tumor core, and edema regions. The baseline model (Original GMMAS) shows the Dice scores of 0.911 ± 0.048, 0.879 ± 0.111, and 0.811 ± 0.139 for the whole tumor, tumor core, and edema regions, respectively.

\begin{table*}[ht]
\caption{Comparison table of segmentation results (mean $\pm$ std)}
\label{table2}
\centering
\renewcommand{\arraystretch}{1.35}
\makebox[\textwidth][c]{ 
    \resizebox{0.85\textwidth}{!}{ 
        \begin{tabular}{>{\arraybackslash}p{3.5cm} c c c }
        \hline
        \textbf{} & \textbf{Whole tumor region} & \textbf{Tumor core region} & \textbf{Edema region} \\
        \hline
        Original GMMAS & 0.911 $\pm$ 0.048 & 0.879 $\pm$ 0.111 & 0.811 $\pm$ 0.139 \\
        + tumor voxels filter & 0.917 $\pm$ 0.042 & 0.887 $\pm$ 0.102 & 0.824 $\pm$ 0.138 \\
        + global feature fusion & 0.933 $\pm$ 0.033 & 0.914 $\pm$ 0.101 & 0.849 $\pm$ 0.115 \\
        + semi-supervised learning & \textbf{0.940 $\pm$ 0.037} & \textbf{0.919 $\pm$ 0.092} & \textbf{0.870 $\pm$ 0.101} \\
        \hline
        \multicolumn{4}{l}{\textit{+: This symbol indicates that new techniques have been added based on the conditions of the previous line.}} 
        \end{tabular}
    }
}
\vspace{-1em}
\end{table*}

\begin{figure*}[t]
    \centering
    \includegraphics[width=\linewidth]{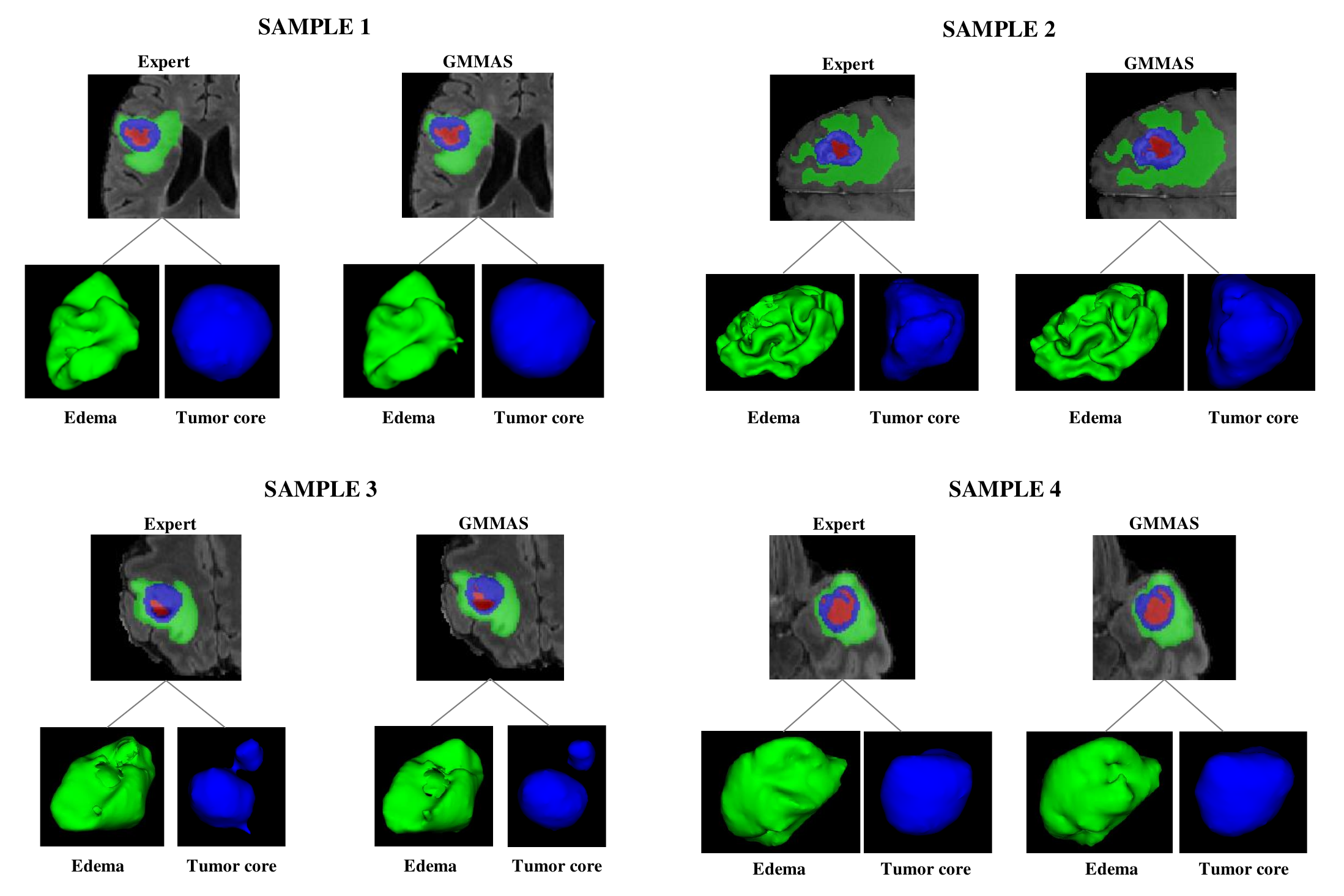}
    \captionsetup{skip=2pt}
    \caption{Comparison of human expert segmentation results (right) and model segmentation results (left) and their respective 3D visualization of edema and tumor core region.}
    \label{fig11}
    \vspace{-1em}
\end{figure*}

The first enhancement to the model is the integration of a tumor voxel filter. This is a post-processing approach that refines segmentation by reducing noise and improving the clarity of tumor boundaries. As shown in Figure~\ref{fig8}, the filter selectively keeps highly likely tumor voxels, improving the model's overall segmentation precision across various regions. Second, the incorporation of a global feature fusion module helps extract more comprehensive MRI features that encapsulate both global and local tumor characteristics, as illustrated in Figure~\ref{fig2}. This strategy enhances the model's ability to recognize and integrate wide-ranging spatial information, ensuring a richer and more robust feature set. As a result, the Dice scores for each region are 0.933±0.033, 0.914±0.101, and 0.849±0.115, respectively.

To further enhance the model’s capability, we designed the two-stage semi-supervised learning method, as illustrated in Figure~\ref{fig10}, to fully exploit the large amount of unlabeled data for classification tasks. Due to the multi-task learning nature of shared representations, the increasing exploration of classification tasks also enhances the segmentation performance by refining the learned shared feature maps. Through the utilization of semi-supervised learning, the Dice scores have been boosted to 0.940 ± 0.037, 0.919 ± 0.092, and 0.870 ± 0.101 for the whole tumor, tumor core, and edema regions, respectively.

Figure~\ref{fig11} shows four cases in which the segmentation capabilities of the GMMAS model were directly compared with those of human experts. In these 2D MRI slices of patients, different colors represent different tumor subregions, where red, blue, and green represent necrosis, enhancing tumors, and edema region. 3D visualizations of edema and tumor core (including necrosis and enhancing tumor region) are shown below the 2D slices. The expert segmentation is shown together with the GMMAS results for comparison. The visual comparison of axial slices and 3D visualizations underscores the precision with which the GMMAS model matches the segmentation of the expert. The 3D visualizations reveal the spatial complexity of the tumor structures, with the GMMAS model still demonstrating a high degree of morphological accuracy, on a par with that of human experts.

\subsubsection{Histological and molecular subtyping of glioma}
For classification tasks, GMMAS' performance also improves with the integration of the global feature fusion module and semi-supervised learning. As demonstrated by the metrics in Table~\ref{table3}, including the Area Under the Curve (AUC), Accuracy (Acc), and F1 score, these methods have shown their effectiveness in enabling GMMAS to better classify glioma histological subtypes and predict biomarkers.

The original GMMAS setup served as a baseline. For histological subtype prediction, it reached an AUC, Acc, and F1 score of 0.962, 0.909, and 0.911, respectively. The application of the global feature fusion module, as illustrated in Figure~\ref{fig2}, enhanced these metrics to 0.975, 0.914, and 0.916. Through this approach, GMMAS provided a deeper understanding of the imaging data by integrating global MRI features that capture wider and more varied patterns across the entire scan, and local features that focus on specific tumor characteristics. This holistic and comprehensive approach enables GMMAS to effectively identify subtle distinctions between different glioma subtypes. Further enhancement was achieved through the two-stage semi-supervised learning, as shown in Figure~\ref{fig10}. The training set was expanded by generating and selecting reliable pseudo-labels through the pretrained GMMAS model, intuitively enabling effective self-training from a broader array of learning examples. Moreover, the second stage of semi-supervised learning, consistency regularization, further enhanced the model’s generalization and stability. These two stages collectively boosted the metrics: AUC to 0.980, Acc to 0.941, and F1 score to 0.943. Similar improvements have also been observed in the biomarker prediction of IDH and the 1p/19q, with their AUC values reaching 0.982 and 0.944, respectively.

\begin{table*}[ht]
\caption{Comparison table of classification results} 
\label{table3}
\centering
\renewcommand{\arraystretch}{1.3}
\fontsize{10.5}{12}\selectfont
\makebox[\textwidth][c]{ 
    \resizebox{0.8\textwidth}{!}{ 
        \begin{tabular*}{\textwidth}{@{\extracolsep{\fill}} p{7cm} c c c c @{}}
        \hline
        \centering\textbf{Histological subtype prediction} & \textbf{AUC} & \textbf{Acc} & \textbf{F1 score} \\
        \hline
        \centering Original GMMAS & 0.962 & 0.909 & 0.911 \\

        \centering+ global feature fusion & 0.970 & 0.925 & 0.926 \\
        
        \centering+ semi-supervised learning & \textbf{0.980} & \textbf{0.941} & \textbf{0.943} \\
        \hline
        \centering\textbf{Biomarker - IDH} & \textbf{AUC} & \textbf{Acc} & \textbf{F1 score} \\
        \hline
        \centering Original GMMAS & 0.964 & 0.905 & 0.842 \\
        
        \centering+ global feature fusion & 0.969 & 0.933 & 0.878 \\
        
        \centering+ semi-supervised learning & \textbf{0.982} & \textbf{0.950} & \textbf{0.917} \\
        \hline
        \centering\textbf{Biomarker - 1p/19q} & \textbf{AUC} & \textbf{Acc} & \textbf{F1 score} \\
        \hline
        \centering Original GMMAS & 0.925 & 0.854 & 0.824 \\
        
        \centering+ global feature fusion & 0.930 & 0.874 & 0.848 \\
        
        \centering+ semi-supervised learning & \textbf{0.944} & \textbf{0.896} & \textbf{0.875} \\
        \hline
        \end{tabular*}
        }
 } 
\vspace{-1em}
\end{table*}

\begin{figure*}[t]
    \centering
    \includegraphics[width=\linewidth]{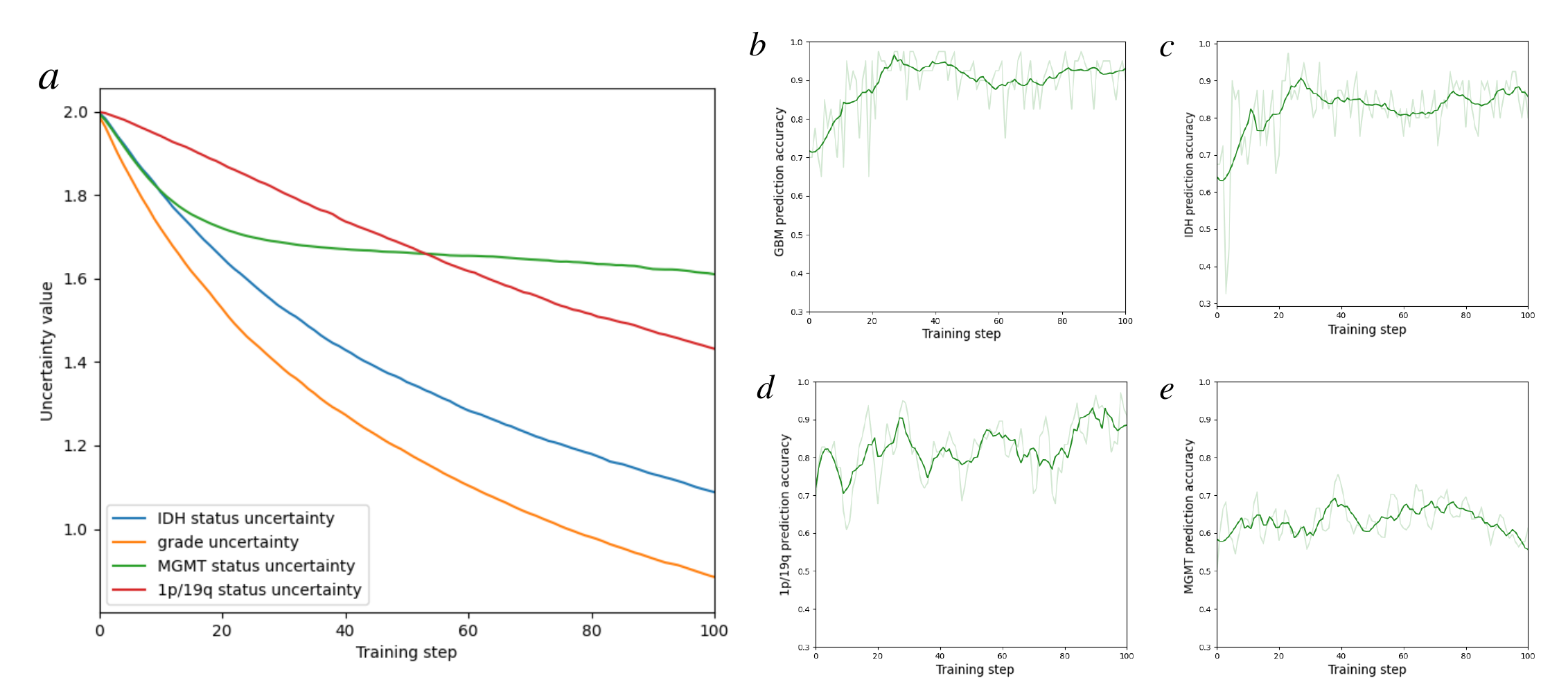}
    \caption{\textbf{Aleatoric uncertainty and accuracy of each task.} (a) Aleatoric uncertainty values during 100 training steps. Classification accuracy curves of (b) glioma grade prediction, (c) IDH mutation status prediction, (d) 1p/19q chromosome disorder prediction, and (e) MGMT status prediction during 100 training steps.}
    \label{fig6}
\end{figure*}

Figure~\ref{fig6} shows that we added the predicted biomarkers 1p/19q and MGMT by extending the multi-classifier module. We retrained only the classifier module and copied the remaining components using the transfer learning method. Figure~\ref{fig6}(a) shows the aleatoric uncertainty values during the training steps, which reflect the correlation between labels and input MRI data; a higher value indicates a higher difficulty when the model has to solve this task. The accuracies of the model on these different tasks reflect their respective prediction complexities. Experimental results confirm that as the aleatoric uncertainty value—which indicates task complexity and difficulty—increases, the prediction accuracy decreases accordingly. The aleatoric uncertainty of each prediction task decreases with an increase in the training steps, proving that the features learned by the network gradually become more representative with model training, which makes the glioma subtypes predictable.

The relatively lower uncertainty and higher prediction accuracy for IDH status and glioma grade imply that these subtypes have more discernible patterns or signatures on MRI scans, making them easier to predict. These patterns can be attributed to the distinct phenotypic characteristics of the gliomas captured by imaging, which are closely linked to the underlying genetic alterations.

Conversely, the higher uncertainty in predicting biomarkers such as MGMT status and 1p/19q status suggests that these tasks are more challenging for the model. This can be attributed to a more subtle manifestation of these genetic alterations in the imaging data, which do not translate as clearly into the MRI features used by the model to make predictions. Further, it can reflect a more complex relationship between these biomarkers and imaging characteristics, requiring the model to capture deeper or nonlinear patterns to improve prediction accuracy.

\subsubsection{Ablation study}

As shown in Table~\ref{table5}, our multi-task model utilizing uncertainty-based task weighting not only outperformed the single-task models in each task but also showed improvements over a plain multi-task learning model, which uses a straightforward summation of various tasks during training. Moreover, we evaluated the traditional ConvNet model that uses convolutional layers instead of the transformer blocks and obtained relatively decreased performance on each task. These results demonstrate that training a model on related tasks simultaneously enriches the training signals and diversifies the learning goals, which in turn cultivates a more robust and adaptive model. The integration of transformer modules also enhances performance by allowing the model to dynamically prioritize relevant features through self-attention mechanisms, optimizing both task-specific and overall model performance.

Table~\ref{table6} compares various semi-supervised learning methods across segmentation and classification tasks. We use the fully-supervised model as a baseline to compare the performance of various semi-supervised learning (SSL) methods. Specifically, we compare FixMatch~\cite{sohn2020fixmatch} and CPS~\cite{chen2021semi}, which represent different approaches in the SSL paradigm. Additionally, we decompose and analyze our two-stage method, demonstrating that combining both stages effectively enhances model performance. Our experiments show that the two-stage approach consistently outperforms the other SSL methods, highlighting the benefits of integrating these two stages to achieve superior results across segmentation and classification tasks.

\begin{figure*}[t]
    \centering
    \includegraphics[width=\linewidth]{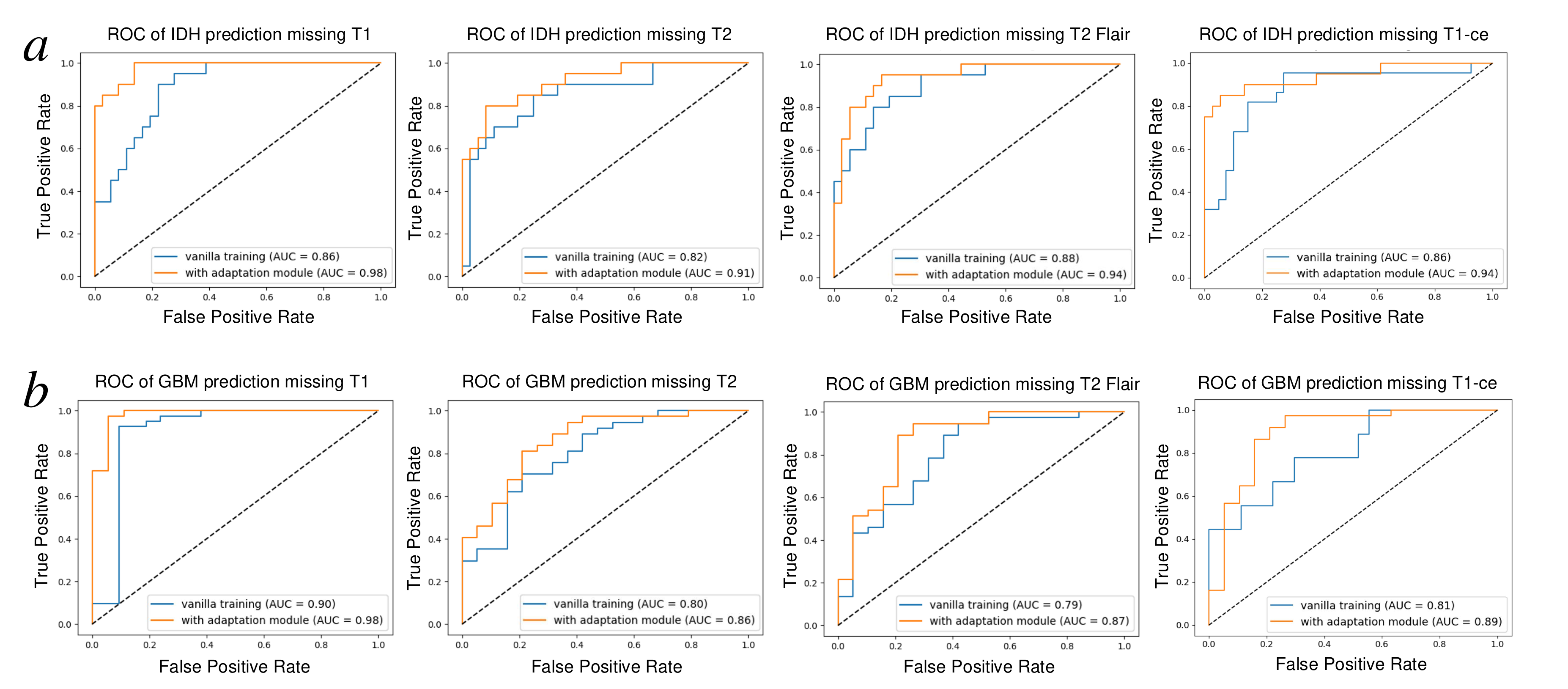}
    \caption{ROC curves of the vanilla training model and the model with an adaptation module while missing different modalities. (a) ROC curves of IDH genotyping. (b) ROC curves of histological subtype prediction.}
    \label{fig12}
\end{figure*}

\subsection{Adaptation ability to modal absence}
Figure~\ref{fig12} shows the ROC curves of the vanilla training model and our enhanced model with an adaptation module designed for cross-modal feature learning. The specialized adaptation module is designed to increase the model's resilience to missing input modalities, allowing it to adapt to incomplete data scenarios.

Figure~\ref{fig12}(a) shows the ROC curves for IDH genotyping, which differentiates between IDH wild type and mutant types. Specifically, when the T1 imaging modality was missing, the AUC increased from 0.86 in the vanilla training model to 0.98 with the adaptation module. For missing T2 data, the AUC improved from 0.82 to 0.91. Similarly, in the absence of T2-Flair imaging, the adaptation-enhanced model achieved an AUC of 0.94, compared to 0.88 with the vanilla training model.

Figure~\ref{fig12}(b) shows the assessment of histological subtypes—low-grade glioma versus glioblastoma—where the adaptation module significantly enhanced model performance in scenarios with missing imaging modalities. For cases lacking T1 imaging, the AUC for the model improved from 0.90 in the vanilla training to 0.98 with the adaptation module. In the absence of T2 data, the AUC increased from 0.80 to 0.86. Similarly, the enhanced model improved the AUC to 0.87 when T2-Flair is absent.

The inclusion of the adaptation module resulted in improvements in AUC values across all scenarios, underscoring the module's effectiveness in handling data incompleteness. By leveraging cross-modal feature learning, a model can extract universal insights from the available data, enhancing its ability to perform consistently under different conditions. This capability of the adaptation module allows the model to compensate for missing data modalities by exploiting the informative content of the remaining data.

Table~\ref{table7} and \ref{table8} evaluate the segmentation Dice scores for the GMMAS model across different imaging modalities to ascertain the impact of missing data on segmentation performance. With full modal input, fully-supervised GMMAS achieves the highest segmentation accuracy for each tumor subregion, showing a Dice score of 0.933 ± 0.033 for the whole tumor region.

When the T2-Flair modality is absent, there is a slight decrease in performance, with Dice scores dropping to 0.916 ± 0.064 for the whole tumor region and 0.821 ± 0.169 for the edema region, while the tumor core region remains relatively stable at 0.910 ± 0.117. This result emphasizes that the T2-Flair modality is crucial for the model to effectively distinguish the edema region from the MRI images. Additionally, this is in line with the expert experience that T2-Flair is essential for detecting and clearly displaying areas of brain edema, including inflammation, infection, ischemic lesions, and other brain diseases associated with edema.

The absence of the T1CE modality results in Dice scores of 0.931 ± 0.044 for the whole tumor and 0.845 ± 0.119 for the edema region, with a more pronounced impact on the tumor core region, dropping to 0.882 ± 0.163. This highlights the significant role of T1CE in delineating the boundary of the tumor core and enhancing tumor visibility. This importance is further supported by the enhanced brightness of the tumor area in T1CE MRI images and the tendency of the contrast agent to accumulate in areas with dense blood vessels or increased blood vessel permeability.

The absence of the T1 modality slightly reduces the model's performance, with the whole tumor region's Dice score dropping to 0.915 ± 0.060 and the edema region’s to 0.827 ± 0.142. The tumor core region remains relatively unaffected with a Dice score of 0.909 ± 0.123. The absence of the T2 modality results in a Dice score of 0.922 ± 0.051 for the whole tumor region; however, the tumor core and edema regions experience decreases to 0.889 ± 0.155 and 0.830 ± 0.151, respectively. This suggests that T2 is integral in helping the model differentiate between the edema and tumor core areas.

In Table~\ref{table8}, the segmentation performance of the GMMAS model under combined modality deletions further highlights the model's robust adaptability to complex clinical scenarios and underscores the efficacy of its adaptation module. When multiple MRI modalities are absent, the model demonstrates a controlled degradation in performance, managing to maintain functional Dice scores for tumor region segmentation despite significant data limitations. This resilience affirms the model's capacity to handle diverse clinical conditions, ensuring reliable segmentation outcomes essential for effective diagnosis and treatment planning.

\begin{table*}[!t]
\caption{Segmentation Dice of Different Modal Absences}
\label{table7}
\centering
\fontsize{10.5}{15}\selectfont
\resizebox{0.85\textwidth}{!}{
\begin{tabular}{>{\centering}p{5cm} p{4cm} p{4cm} p{4cm}}
\hline
\multirow{2}{*}{\textbf{Methods}} & \multicolumn{3}{c}{\textbf{Segmentation} (Dice Mean $\pm$ Std)}\\ \cline{2-4}
                        & Whole Tumor Region & Tumor Core Region & Edema Region \\ \hline
\textbf{Full Modal Input}         & \textbf{0.933 $\pm$ 0.033}   & \textbf{0.914 $\pm$ 0.101}  & \textbf{0.849 $\pm$ 0.115} \\ 
Missing T2-Flair         & 0.916 $\pm$ 0.064   & \textbf{0.910 $\pm$ 0.117}  & 0.821 $\pm$ 0.169 \\ 
Missing T1CE            & \textbf{0.931 $\pm$ 0.044}   & 0.882 $\pm$ 0.163  & \textbf{0.845 $\pm$ 0.119} \\ 
Missing T1               & 0.915 $\pm$ 0.060   & 0.909 $\pm$ 0.123  & 0.827 $\pm$ 0.142 \\ 
Missing T2               & 0.922 $\pm$ 0.051   & 0.889 $\pm$ 0.155  & 0.830 $\pm$ 0.151 \\ \hline
\end{tabular}
}
\end{table*}

\begin{table*}[!t]
\caption{Segmentation Dice of Two Different Modal Absences}
\label{table8}
\centering
\fontsize{10.5}{15}\selectfont
\resizebox{0.85\textwidth}{!}{
\begin{tabular}{>{\centering}p{5cm} p{4cm} p{4cm} p{4cm}}
\hline
\multirow{2}{*}{\textbf{Methods}} & \multicolumn{3}{c}{\textbf{Segmentation} (Dice Mean $\pm$ Std)} \\ \cline{2-4}
                        & Whole Tumor Region & Tumor Core Region & Edema Region \\ \hline
Missing T2-Flair \& T1CE & 0.903 $\pm$ 0.109 & 0.873 $\pm$ 0.184 & 0.812 $\pm$ 0.188 \\ 
Missing T2-Flair \& T1    & 0.899 $\pm$ 0.110 & \textbf{0.881 $\pm$ 0.171} & 0.804 $\pm$ 0.195 \\ 
Missing T2-Flair \& T2    & 0.884 $\pm$ 0.109 & 0.879 $\pm$ 0.176 & 0.806 $\pm$ 0.191 \\ 
Missing T1CE \& T1       & 0.893 $\pm$ 0.101 & 0.870 $\pm$ 0.183 & \textbf{0.819 $\pm$ 0.179} \\ 
Missing T1CE \& T2       & \textbf{0.909 $\pm$ 0.092} & 0.869 $\pm$ 0.187 & 0.818 $\pm$ 0.184 \\ 
Missing T1 \& T2          & 0.883 $\pm$ 0.113 & 0.874 $\pm$ 0.180 & 0.807 $\pm$ 0.193 \\ \hline
\end{tabular}
}
\end{table*}

\begin{figure*}[t]
    \centering
    \includegraphics[width=\linewidth]{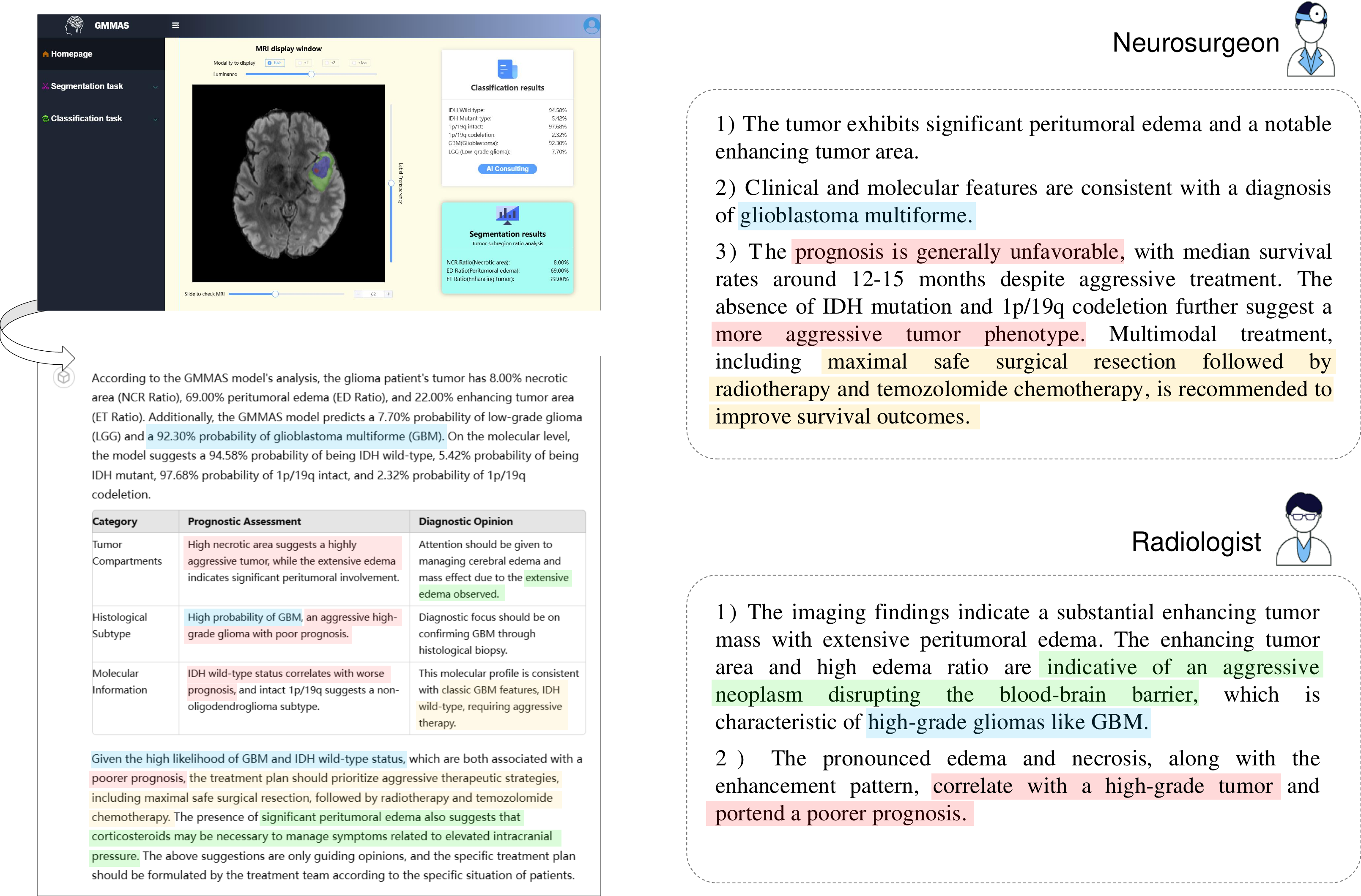}
    \caption{\textbf{GMMAS platform and comparison of GMMAS results with expert evaluations from a neurosurgeon and a radiologist.} The GMMAS analysis results include MRI segmentation results indicating proportions of subregions within tumors and classification results detailing the molecular and histological characteristics of the glioma. The combined results are fed into GMMAS-GPT that provides a prognostic assessment and diagnostic opinion about this case. Colored text highlights semantic agreement between GMMAS-generated assessments and expert opinions, demonstrating the model's accuracy in mirroring professional diagnostic and prognostic evaluations.}
    \label{fig13}
    \vspace{-1.2em}
\end{figure*}

\subsection{Platform establishment and introducing GMMAS-GPT }
Finally, we created a platform, as shown in Figure~\ref{fig13}, to visualize our model and provide a website for doctors and patients. We present a video in the supplementary data to show the main functional modules. This platform serves as a user-friendly interface that combines the robust GMMAS model for tumor segmentation and subtyping with the advanced natural language processing capabilities of a large language model for generating prognostic assessments and therapeutic recommendations.

On uploading the MRI scans to the platform, the GMMAS model performed simultaneous segmentation and classification of the MRI scan data, providing real-time results. Users receive immediate visual feedback on the segmentation results, which are then quantitatively detailed with the proportion of each subregion of the tumor: necrotic area, edema, and enhancing tumor area. Concurrently, classification was performed to determine the histological, genetic, and molecular characteristics of the tumor, including glioma grade, IDH mutation status, and 1p/19q codeletion status. The classification results are shown with their prediction confidence, which is well-calibrated and reflects their real possibilities.

In the prognostic assessment section, these analysis results are fed into the GMMAS-GPT language model, which is a GPT-based agent with a knowledge base including the “Guidelines for diagnosis of glioma,” “The 2021 WHO Classification,” and a series of professional medical documents. The language model automatically takes the analysis results as prompts and generates a comprehensive report in a chart format. The report includes a prognostic assessment indicating tumor behavior and expected prognosis, as well as a diagnostic opinion correlated with glioma subtypes. For example, a high percentage of enhanced tumor area can indicate a more aggressive or advanced stage of the tumor, influencing the urgency and type of medical intervention required. While providing a professional response, GMMAS-GPT will also inform the user at the end of the response that ``The above recommendations are guidelines only, and specific treatment plans should be developed by the treatment team on a patient-specific basis."

On the right of Figure~\ref{fig13}, the corresponding evaluations from medical experts align closely with the AI-generated conclusions, as indicated by the color-coded text matching. This alignment showcases the GMMAS' capability to replicate expert-level medical assessments, potentially augmenting clinical decision-making processes in glioma management.

\section{Future Work and Conclusions}
In this study, we introduced GMMAS to improve the auxiliary diagnosis of glioma through MRI analysis. GMMAS offers a novel approach for enhancing diagnostic precision by integrating multi-task semi-supervised learning. The adaptation ability of GMMAS to various imaging modalities is a notable feature designed to provide flexibility in clinical settings where certain MRI modalities may be unavailable. A visual and user-friendly platform, along with the introduction of GMMAS-GPT to generate personalized prognostic evaluations, provides a more comprehensive diagnostic tool based on AI technologies. The potential for expanding GMMAS' application beyond gliomas and integrating it into clinical workflows promises to greatly facilitate clinical decision-making processes by creating a more efficient multicenter, multimodal, and multifunctional model.

Future studies will focus on collecting a broader array of patient samples from multiple centers to enhance the diversity and representativeness of the study population. Further, we also aim to explore the extension of the GMMAS diagnostic capabilities to expand the scope of the system for including other types of CNS tumors, such as meningiomas and ependymomas. Another area for future development is incorporating multimodal inputs, including whole-slice imaging from tissue biopsies. This enhancement can provide a richer dataset for prognostic assessment, leveraging the strengths of various imaging techniques to offer a more detailed tumor characterization. Further, the GMMAS-GPT component will be refined for incorporating a wider range of expert knowledge with the aim of supporting the system in providing more professional and personalized diagnostic opinions. The pursuit of these future directions is essential for realizing the full potential of GMMAS in clinical practice.

\section*{Declaration of Competing Interest}
The authors declare that they have no known competing financial interests or personal relationships that could have appeared to influence the work reported in this paper.

\section*{Acknowledgment}
This work was supported in part by the High Performance Computing Center of Central South University. This work was supported by grants from the National Natural Science Foundation of China (82073096) and Graduate Research and Innovation Projects of Central South University (1053320221550).

\section*{Supplementary material}
There is a supplementary video of the visual platform for this research.

{
\bibliographystyle{IEEEtran}
\bibliography{main}

\begin{thebibliography}{10}
\providecommand{\url}[1]{#1}
\csname url@samestyle\endcsname
\providecommand{\newblock}{\relax}
\providecommand{\bibinfo}[2]{#2}
\providecommand{\BIBentrySTDinterwordspacing}{\spaceskip=0pt\relax}
\providecommand{\BIBentryALTinterwordstretchfactor}{4}
\providecommand{\BIBentryALTinterwordspacing}{\spaceskip=\fontdimen2\font plus
\BIBentryALTinterwordstretchfactor\fontdimen3\font minus \fontdimen4\font\relax}
\providecommand{\BIBforeignlanguage}[2]{{%
\expandafter\ifx\csname l@#1\endcsname\relax
\typeout{** WARNING: IEEEtran.bst: No hyphenation pattern has been}%
\typeout{** loaded for the language `#1'. Using the pattern for}%
\typeout{** the default language instead.}%
\else
\language=\csname l@#1\endcsname
\fi
#2}}
\providecommand{\BIBdecl}{\relax}
\BIBdecl

\bibitem{louis20162016}
D.~N. Louis, A.~Perry, G.~Reifenberger, A.~Von~Deimling, D.~Figarella-Branger, W.~K. Cavenee, H.~Ohgaki, O.~D. Wiestler, P.~Kleihues, and D.~W. Ellison, ``The 2016 world health organization classification of tumors of the central nervous system: a summary,'' \emph{Acta neuropathologica}, vol. 131, pp. 803--820, 2016.

\bibitem{louis20212021}
D.~N. Louis, A.~Perry, P.~Wesseling, D.~J. Brat, I.~A. Cree, D.~Figarella-Branger, C.~Hawkins, H.~Ng, S.~M. Pfister, G.~Reifenberger \emph{et~al.}, ``The 2021 who classification of tumors of the central nervous system: a summary,'' \emph{Neuro-oncology}, vol.~23, no.~8, pp. 1231--1251, 2021.

\bibitem{yang2022glioma}
K.~Yang, Z.~Wu, H.~Zhang, N.~Zhang, W.~Wu, Z.~Wang, Z.~Dai, X.~Zhang, L.~Zhang, Y.~Peng \emph{et~al.}, ``Glioma targeted therapy: insight into future of molecular approaches,'' \emph{Molecular Cancer}, vol.~21, no.~1, p.~39, 2022.

\bibitem{lv2024insight}
Q.~Lv, Y.~Liu, Y.~Sun, and M.~Wu, ``Insight into deep learning for glioma idh medical image analysis: A systematic review,'' \emph{Medicine}, vol. 103, no.~7, p. e37150, 2024.

\bibitem{pirozzi2021implications}
C.~J. Pirozzi and H.~Yan, ``The implications of idh mutations for cancer development and therapy,'' \emph{Nature reviews Clinical oncology}, vol.~18, no.~10, pp. 645--661, 2021.

\bibitem{lv2024predictive}
Q.~Lv, Z.~Zhang, H.~Fu, D.~Li, Y.~Liu, Y.~Sun, and M.~Wu, ``Predictive panel for immunotherapy in low-grade glioma,'' \emph{World Neurosurgery}, vol. 183, pp. e825--e837, 2024.

\bibitem{liu2025survey}
Y.~Liu, X.~Cao, T.~Chen, Y.~Jiang, J.~You, M.~Wu, X.~Wang, M.~Feng, Y.~Jin, and J.~Chen, ``A survey of embodied ai in healthcare: Techniques, applications, and opportunities,'' \emph{arXiv preprint arXiv:2501.07468}, 2025.

\bibitem{park2018prediction}
Y.~Park, K.~Han, S.~Ahn, S.~Bae, Y.~Choi, J.~Chang, S.~Kim, S.-G. Kang, and S.-K. Lee, ``Prediction of idh1-mutation and 1p/19q-codeletion status using preoperative mr imaging phenotypes in lower grade gliomas,'' \emph{American Journal of Neuroradiology}, vol.~39, no.~1, pp. 37--42, 2018.

\bibitem{cheng2022fully}
J.~Cheng, J.~Liu, H.~Kuang, and J.~Wang, ``A fully automated multimodal mri-based multi-task learning for glioma segmentation and idh genotyping,'' \emph{IEEE Transactions on Medical Imaging}, vol.~41, no.~6, pp. 1520--1532, 2022.

\bibitem{hao2023predicting}
X.~Hao, H.~Xu, N.~Zhao, T.~Yu, T.~Hamalainen, and F.~Cong, ``Predicting pathological complete response based on weakly and semi-supervised joint learning from breast cancer mri,'' in \emph{2023 45th Annual International Conference of the IEEE Engineering in Medicine \& Biology Society (EMBC)}.\hskip 1em plus 0.5em minus 0.4em\relax IEEE, 2023, pp. 1--4.

\bibitem{parmar2023generalizable}
C.~Parmar, A.~J. Ramon, N.~L. Stone, S.~Triantos, J.~Greshock, and K.~Standish, ``Generalizable fgfr prediction across tumor types using self supervised learning.'' 2023.

\bibitem{chen2021exploring}
X.~Chen and K.~He, ``Exploring simple siamese representation learning,'' in \emph{Proceedings of the IEEE/CVF conference on computer vision and pattern recognition}, 2021, pp. 15\,750--15\,758.

\bibitem{kendall2018multi}
A.~Kendall, Y.~Gal, and R.~Cipolla, ``Multi-task learning using uncertainty to weigh losses for scene geometry and semantics,'' in \emph{Proceedings of the IEEE conference on computer vision and pattern recognition}, 2018, pp. 7482--7491.

\bibitem{liu2023deep}
Y.~Liu and M.~Wu, ``Deep learning in precision medicine and focus on glioma,'' \emph{Bioengineering \& Translational Medicine}, vol.~8, no.~5, p. e10553, 2023.

\bibitem{yamashita2018convolutional}
R.~Yamashita, M.~Nishio, R.~K.~G. Do, and K.~Togashi, ``Convolutional neural networks: an overview and application in radiology,'' \emph{Insights into imaging}, vol.~9, pp. 611--629, 2018.

\bibitem{mazurowski2019deep}
M.~A. Mazurowski, M.~Buda, A.~Saha, and M.~R. Bashir, ``Deep learning in radiology: An overview of the concepts and a survey of the state of the art with focus on mri,'' \emph{Journal of magnetic resonance imaging}, vol.~49, no.~4, pp. 939--954, 2019.

\bibitem{wu2019radiomics}
S.~Wu, J.~Meng, Q.~Yu, P.~Li, and S.~Fu, ``Radiomics-based machine learning methods for isocitrate dehydrogenase genotype prediction of diffuse gliomas,'' \emph{Journal of cancer research and clinical oncology}, vol. 145, pp. 543--550, 2019.

\bibitem{ren2019noninvasive}
Y.~Ren, X.~Zhang, W.~Rui, H.~Pang, T.~Qiu, J.~Wang, Q.~Xie, T.~Jin, H.~Zhang, H.~Chen \emph{et~al.}, ``Noninvasive prediction of idh1 mutation and atrx expression loss in low-grade gliomas using multiparametric mr radiomic features,'' \emph{Journal of Magnetic Resonance Imaging}, vol.~49, no.~3, pp. 808--817, 2019.

\bibitem{corso2008efficient}
J.~J. Corso, E.~Sharon, S.~Dube, S.~El-Saden, U.~Sinha, and A.~Yuille, ``Efficient multilevel brain tumor segmentation with integrated bayesian model classification,'' \emph{IEEE transactions on medical imaging}, vol.~27, no.~5, pp. 629--640, 2008.

\bibitem{parmar2015radiomic}
C.~Parmar, R.~T. Leijenaar, P.~Grossmann, E.~Rios~Velazquez, J.~Bussink, D.~Rietveld, M.~M. Rietbergen, B.~Haibe-Kains, P.~Lambin, and H.~J. Aerts, ``Radiomic feature clusters and prognostic signatures specific for lung and head \& neck cancer,'' \emph{Scientific reports}, vol.~5, no.~1, p. 11044, 2015.

\bibitem{waghere2024robust}
S.~S. Waghere and J.~P. Shinde, ``A robust classification of brain tumor disease in mri using twin-attention based dense convolutional auto-encoder,'' \emph{Biomedical Signal Processing and Control}, vol.~92, p. 106088, 2024.

\bibitem{li2024multi}
S.~Li and S.~Guo, ``Multi-task parallel with feature sharing integrated 3d u-nets for glioma segmentation,'' \emph{Biomedical Signal Processing and Control}, vol.~93, p. 106178, 2024.

\bibitem{matsui2020prediction}
Y.~Matsui, T.~Maruyama, M.~Nitta, T.~Saito, S.~Tsuzuki, M.~Tamura, K.~Kusuda, Y.~Fukuya, H.~Asano, T.~Kawamata \emph{et~al.}, ``Prediction of lower-grade glioma molecular subtypes using deep learning,'' \emph{Journal of neuro-oncology}, vol. 146, pp. 321--327, 2020.

\bibitem{tang2020deep}
Z.~Tang, Y.~Xu, L.~Jin, A.~Aibaidula, J.~Lu, Z.~Jiao, J.~Wu, H.~Zhang, and D.~Shen, ``Deep learning of imaging phenotype and genotype for predicting overall survival time of glioblastoma patients,'' \emph{IEEE transactions on medical imaging}, vol.~39, no.~6, pp. 2100--2109, 2020.

\bibitem{vaswani2017attention}
A.~Vaswani, ``Attention is all you need,'' \emph{Advances in Neural Information Processing Systems}, 2017.

\bibitem{gillot2022automatic}
M.~Gillot, B.~Baquero, C.~Le, R.~Deleat-Besson, J.~Bianchi, A.~Ruellas, M.~Gurgel, M.~Yatabe, N.~Al~Turkestani, K.~Najarian \emph{et~al.}, ``Automatic multi-anatomical skull structure segmentation of cone-beam computed tomography scans using 3d unetr,'' \emph{PLoS One}, vol.~17, no.~10, p. e0275033, 2022.

\bibitem{esteva2017dermatologist}
A.~Esteva, B.~Kuprel, R.~A. Novoa, J.~Ko, S.~M. Swetter, H.~M. Blau, and S.~Thrun, ``Dermatologist-level classification of skin cancer with deep neural networks,'' \emph{nature}, vol. 542, no. 7639, pp. 115--118, 2017.

\bibitem{liu2019artificial}
Y.~Liu, T.~Kohlberger, M.~Norouzi, G.~E. Dahl, J.~L. Smith, A.~Mohtashamian, N.~Olson, L.~H. Peng, J.~D. Hipp, and M.~C. Stumpe, ``Artificial intelligence--based breast cancer nodal metastasis detection: insights into the black box for pathologists,'' \emph{Archives of pathology \& laboratory medicine}, vol. 143, no.~7, pp. 859--868, 2019.

\bibitem{keenan2022deeplensnet}
T.~D. Keenan, Q.~Chen, E.~Agr{\'o}n, Y.-C. Tham, J.~H.~L. Goh, X.~Lei, Y.~P. Ng, Y.~Liu, X.~Xu, C.-Y. Cheng \emph{et~al.}, ``Deeplensnet: deep learning automated diagnosis and quantitative classification of cataract type and severity,'' \emph{Ophthalmology}, vol. 129, no.~5, pp. 571--584, 2022.

\bibitem{shorten2019survey}
C.~Shorten and T.~M. Khoshgoftaar, ``A survey on image data augmentation for deep learning,'' \emph{Journal of big data}, vol.~6, no.~1, pp. 1--48, 2019.

\bibitem{yun2019cutmix}
S.~Yun, D.~Han, S.~J. Oh, S.~Chun, J.~Choe, and Y.~Yoo, ``Cutmix: Regularization strategy to train strong classifiers with localizable features,'' in \emph{Proceedings of the IEEE/CVF international conference on computer vision}, 2019, pp. 6023--6032.

\bibitem{falk2019u}
T.~Falk, D.~Mai, R.~Bensch, {\"O}.~{\c{C}}i{\c{c}}ek, A.~Abdulkadir, Y.~Marrakchi, A.~B{\"o}hm, J.~Deubner, Z.~J{\"a}ckel, K.~Seiwald \emph{et~al.}, ``U-net: deep learning for cell counting, detection, and morphometry,'' \emph{Nature methods}, vol.~16, no.~1, pp. 67--70, 2019.

\bibitem{han2022survey}
K.~Han, Y.~Wang, H.~Chen, X.~Chen, J.~Guo, Z.~Liu, Y.~Tang, A.~Xiao, C.~Xu, Y.~Xu \emph{et~al.}, ``A survey on vision transformer,'' \emph{IEEE transactions on pattern analysis and machine intelligence}, vol.~45, no.~1, pp. 87--110, 2022.

\bibitem{kendall2017uncertainties}
A.~Kendall and Y.~Gal, ``What uncertainties do we need in bayesian deep learning for computer vision?'' \emph{Advances in neural information processing systems}, vol.~30, 2017.

\bibitem{sharma2019missing}
A.~Sharma and G.~Hamarneh, ``Missing mri pulse sequence synthesis using multi-modal generative adversarial network,'' \emph{IEEE transactions on medical imaging}, vol.~39, no.~4, pp. 1170--1183, 2019.

\bibitem{zhang2021self}
L.~Zhang, C.~Bao, and K.~Ma, ``Self-distillation: Towards efficient and compact neural networks,'' \emph{IEEE Transactions on Pattern Analysis and Machine Intelligence}, vol.~44, no.~8, pp. 4388--4403, 2021.

\bibitem{luo2021learning}
Y.~Luo, J.~L{\"u}, X.~Jiang, and B.~Zhang, ``Learning from architectural redundancy: Enhanced deep supervision in deep multipath encoder--decoder networks,'' \emph{IEEE Transactions on Neural Networks and Learning Systems}, vol.~33, no.~9, pp. 4271--4284, 2021.

\bibitem{van2016calibration}
B.~Van~Calster, D.~Nieboer, Y.~Vergouwe, B.~De~Cock, M.~J. Pencina, and E.~W. Steyerberg, ``A calibration hierarchy for risk models was defined: from utopia to empirical data,'' \emph{Journal of clinical epidemiology}, vol.~74, pp. 167--176, 2016.

\bibitem{berthelot2019remixmatch}
D.~Berthelot, N.~Carlini, E.~D. Cubuk, A.~Kurakin, K.~Sohn, H.~Zhang, and C.~Raffel, ``Remixmatch: Semi-supervised learning with distribution alignment and augmentation anchoring,'' \emph{arXiv preprint arXiv:1911.09785}, 2019.

\bibitem{yang2025unimatch}
L.~Yang, Z.~Zhao, and H.~Zhao, ``Unimatch v2: Pushing the limit of semi-supervised semantic segmentation,'' \emph{IEEE Transactions on Pattern Analysis and Machine Intelligence}, 2025.

\bibitem{sohn2020fixmatch}
K.~Sohn, D.~Berthelot, N.~Carlini, Z.~Zhang, H.~Zhang, C.~A. Raffel, E.~D. Cubuk, A.~Kurakin, and C.-L. Li, ``Fixmatch: Simplifying semi-supervised learning with consistency and confidence,'' \emph{Advances in neural information processing systems}, vol.~33, pp. 596--608, 2020.

\bibitem{chen2021semi}
X.~Chen, Y.~Yuan, G.~Zeng, and J.~Wang, ``Semi-supervised semantic segmentation with cross pseudo supervision,'' in \emph{Proceedings of the IEEE/CVF conference on computer vision and pattern recognition}, 2021, pp. 2613--2622.

\end{thebibliography}
}

\clearpage
\onecolumn
\section{Appendix}
\renewcommand{\thefigure}{A\arabic{figure}}
\renewcommand{\thetable}{A\arabic{table}}

\begin{figure}[!ht] 
    \centering
    \includegraphics[width=\linewidth]{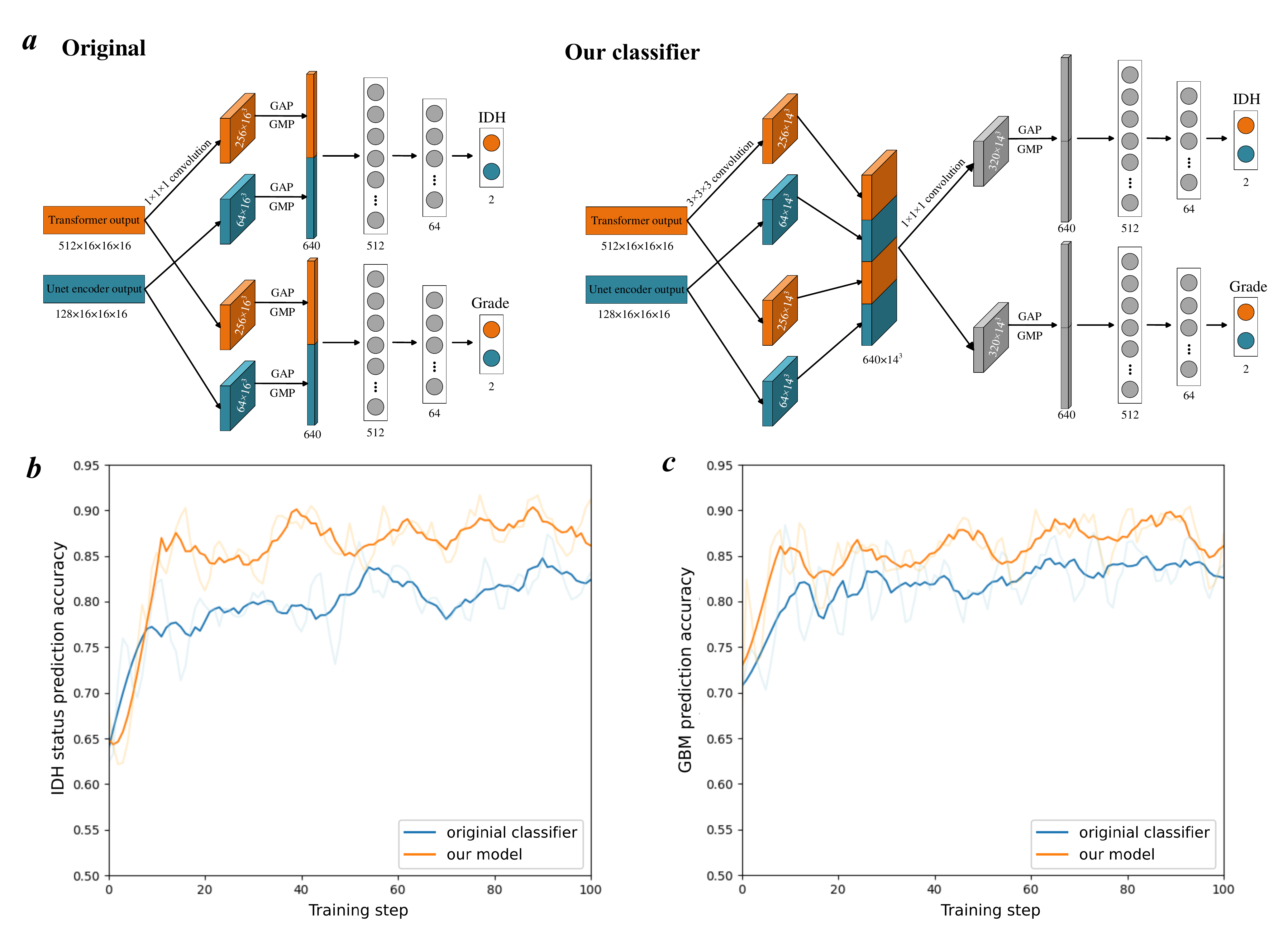}
    \caption{\textbf{Enhanced multi-task classifier and improvements during the training process. }(a) Comparison diagram of the original multi-task classifier and the classifier module of the new model. Classification accuracy curves of (b) IDH mutation status prediction and (c) glioma grade prediction during 100 training steps based on the BraTS 2020 dataset.}
    \label{A1}
\end{figure}

As shown in Figure~\ref{A1}(a), instead of simply stacking the classifier branches, we designed a part of ``Summarize and Separate" to help the model learn more about the cross information between tasks. The original model processed inputs using 1 × 1 × 1 convolutional layers. Both streams were subjected to GAP and GMP (Murray et al., 2014) and concatenated into a dense layer, followed by a classification layer for predicting the IDH mutation status and glioma grade. The GAP captures the average presence of features across the entire image, whereas GMP focuses on the most prominent features. In contrast, our model introduces an architecture that incorporates a ``Summarize and Separate" trick. This architecture ensures that the intertwined biological and morphological information relevant to both the IDH mutation status and glioma grading is adequately captured and utilized before drawing predictions.

Figure~\ref{A1}(b) and (c) depict the performance comparison between the original classifier and our model over 100 training steps for two outputs: IDH status and GBM prediction accuracies. Our model outperformed the original classifier throughout the training steps in both tasks, as demonstrated by the higher accuracy curves shown in the figures. The improved architecture better exploits the inherent correlation between the prediction tasks for the IDH mutation status and glioma subtypes. Using a shared feature-extraction base, our model leverages this correlation to obtain more accurate predictions.

\clearpage
\begin{figure*} 
    \centering
    \includegraphics[width=\linewidth]{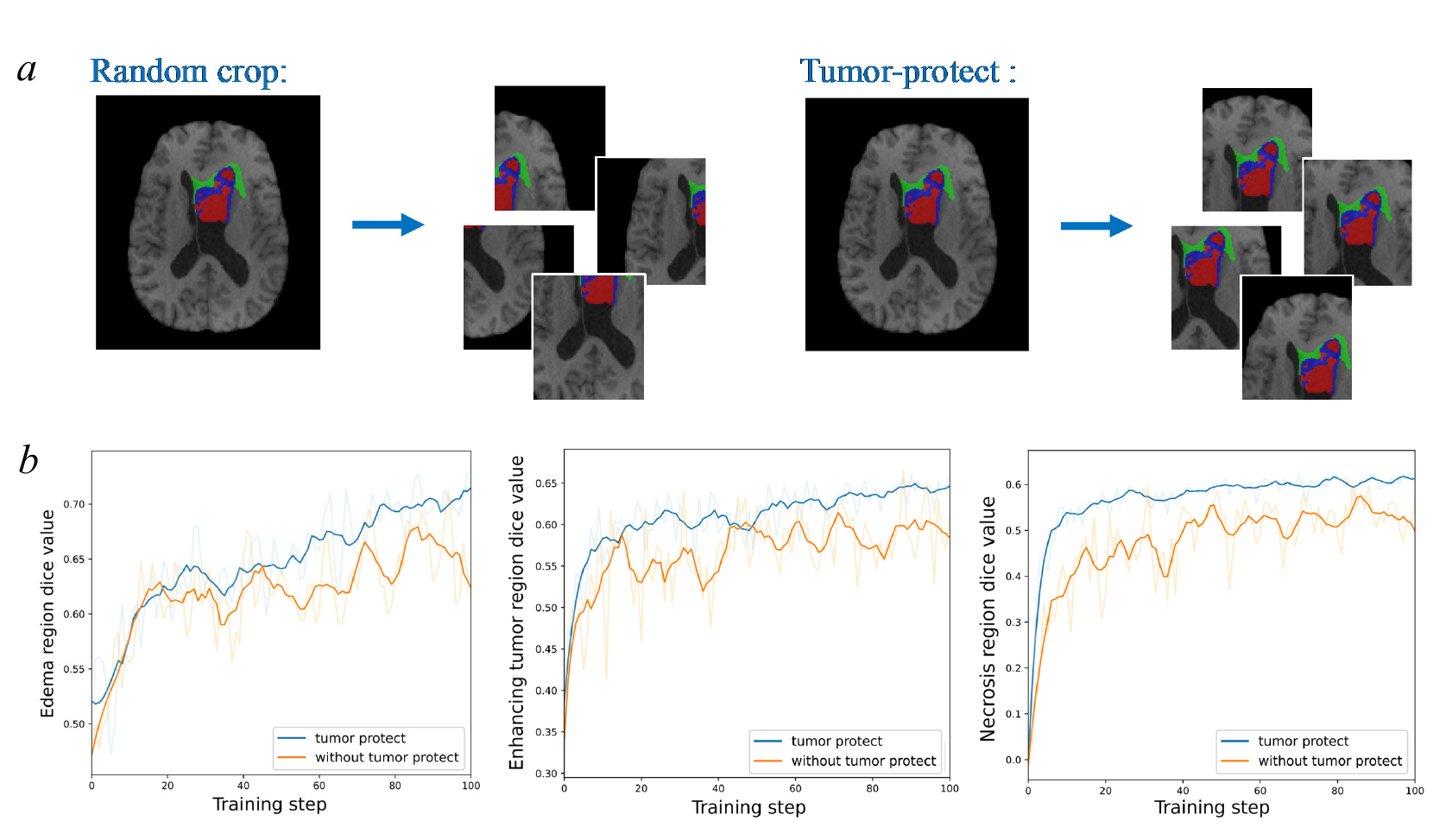}
    \caption{\textbf{Enhanced data augmentation method and improvements to segmentation.} (a) Tumor-protect-crop method. (b) Segmentation Dice coefficients of every tumor subregion during 100 training steps based on the BraTS 2020 dataset.}
    \label{A2}
\end{figure*}

The traditional random crop method does not consider the whole tumor region, which is required by radiologists. This region contains the most important information about glioma patients. As shown in Figure~\ref{A2}(a), the tumor-protect crop method was designed to prioritize the preservation of tumor regions during the cropping process, ensuring that these critical areas are consistently represented in the training batches. Figure~\ref{A2}(b) shows that there is an enhancement in the segmentation accuracy for all subregions of the tumor through employing the tumor-protect crop method. This improvement helped reduce the computational burden by decreasing the image size, enriched the diversity of the samples, and automatically selected regions of interest and focused on useful features contained in MRI.
\clearpage

\begin{table}
\caption{Ablation studies of multi-task learning and transformer blocks}
\label{table5}
\centering
\fontsize{6}{8.5}\selectfont
\renewcommand{\arraystretch}{1.3}
\resizebox{\textwidth}{!}{
\begin{tabular}{>{\raggedright}m{3cm} c c c c c c}
\hline
\multirow{2}{*}{Methods} & \multicolumn{3}{c}{Segmentation (Dice mean $\pm$ std)} & \multicolumn{3}{c}{Classifications} \\ \cline{2-7} 
                        & Whole tumor & Tumor core & Edema & AUC & Acc & F1-score \\ \hline
Single segmentation     & 0.913 $\pm$ 0.089 & 0.897 $\pm$ 0.112 & 0.828 $\pm$ 0.159 & --  & --  & -- \\ 
Single classification   & --  & --  & --  & 0.928 & 0.873 & 0.804 \\ 
Plain multi-task       & 0.917 $\pm$ 0.069 & 0.890 $\pm$ 0.137 & 0.838 $\pm$ 0.142 & 0.931 & 0.898 & 0.855 \\ 
\textbf{Multi-Task}              & \textbf{0.940 $\pm$ 0.037} & \textbf{0.919 $\pm$ 0.092} & \textbf{0.870 $\pm$ 0.101} & \textbf{0.969} & \textbf{0.929} & \textbf{0.912} \\ \hline
ConvNet                & 0.920 $\pm$ 0.057 & 0.877 $\pm$ 0.143 & 0.842 $\pm$ 0.141 & 0.915 & 0.887 & 0.812 \\ \hline
\end{tabular}
}
\end{table}

\begin{table}
\caption{Comparison of different semi-supervised learning methods}
\label{table6}
\centering
\tiny
\renewcommand{\arraystretch}{1.3}
\resizebox{\textwidth}{!}{
\begin{tabular}{>{\raggedright}m{2cm} c c c c c c}
\hline
\multirow{2}{*}{\textbf{Methods}} & \multicolumn{3}{c}{\textbf{Segmentation} (Dice mean $\pm$ std)} & \multicolumn{3}{c}{\textbf{Classifications}} \\ \cline{2-7} 
                        & Whole tumor & Tumor core & Edema & AUC & Acc & F1-score \\ \hline
Fully-supervised        & 0.933 $\pm$ 0.033 & 0.914 $\pm$ 0.101 & 0.849 $\pm$ 0.115 & 0.956 & 0.911 & 0.884 \\ \hline
FixMatch                & 0.935 $\pm$ 0.027 & 0.915 $\pm$ 0.098 & 0.857 $\pm$ 0.110 & 0.958 & 0.916 & 0.898 \\ 
CPS                     & 0.935 $\pm$ 0.038 & 0.916 $\pm$ 0.095 & 0.859 $\pm$ 0.103 & 0.962 & 0.915 & 0.895 \\ 
Stage I-only            & 0.937 $\pm$ 0.024 & 0.916 $\pm$ 0.088 & 0.863 $\pm$ 0.096 & 0.966 & 0.924 & 0.905 \\ 
Stage II-only           & 0.939 $\pm$ 0.041 & 0.917 $\pm$ 0.096 & 0.865 $\pm$ 0.105 & 0.965 & 0.920 & 0.904 \\ 
\textbf{Two-stage SSL }          & \textbf{0.940 $\pm$ 0.037} & \textbf{0.919 $\pm$ 0.092} & \textbf{0.870 $\pm$ 0.101} & \textbf{0.969} & \textbf{0.929} & \textbf{0.912} \\ \hline
\end{tabular}
}
\end{table}

\begin{algorithm}
\caption{Automatic post-processing of the tumor filter}
\label{algA1}
\begin{algorithmic}[1]
\State \textbf{Input:} 3D MRI segmentation result labels Image $I$, where zero pixels represent non-tumor region.
\State Define post\_crop($I$), clip $I$ according to the max coordinates of non-zero pixels, and obtain its proportion $D$ of non-zero pixels within the clipped $I$.
\State Define find\_center($I$) for each non-zero pixel $P(x, y, z)$ and add up the adjacent non-zero pixel value to $P$.
\State Define create\_mask($I$), spread max value of $I$ to the neighboring non-zero pixels, and build mask $M$.
\If{$D < 0.1$}
    \State $C_{img} \gets$ find\_center($I$)
    \State $M \gets$ create\_mask($C_{img}$)
    \State $Threshold \gets \frac{| \{P \mid P \in I \land M[P] = \max(C_{img}) \} |}{| \{P \mid P \in I \land P \neq 0 \} |} < 0.6$
    \If{not Threshold}
        \State $I\{ P \mid M[P] \neq \max(C_{img}) \} \gets 0$
    \Else
        \While{Threshold}
            \State $a: C_{img}\{P \mid M[P] = \max(C_{img}) \} \gets 0$
            \State $b: M \gets$ create\_mask($C_{img}$)
            \State $c: I\{ P \mid M[P] \neq \max(C_{img}) \} \gets 0$
        \EndWhile
    \EndIf
\EndIf
\State $O \gets I$
\State \textbf{Return} $O$
\end{algorithmic}
\end{algorithm}

\end{document}